\documentclass[twocolumn,prl,amsmath,amssymb,showpacs,superscriptaddress]{revtex4-1}
\usepackage{epsf}      
\usepackage{graphicx}
\usepackage{color}
\usepackage{gensymb}
\usepackage{amsmath}
\usepackage{multirow}

\begin{document}

\title {Spin dynamics and unconventional magnetism in insulating La$_{(1-2x)}$Sr$_{2x}$Co$_{(1-x)}$Nb$_{x}$O$_3$}

\author{Rishabh Shukla}
\affiliation{Department of Physics, Indian Institute of Technology Delhi, Hauz Khas, New Delhi-110016, India}
\author{Anil Jain}
\affiliation{Solid State Physics Division, Bhabha Atomic Research Centre, Mumbai-400085, India}
\author{M. Miryala}
\affiliation{Shibaura Institute of Technology, Toyosu campus, Koto-ku Tokyo, Japan}
\author{M. Murakami}
\affiliation{Shibaura Institute of Technology, Toyosu campus, Koto-ku Tokyo, Japan}
\author{K. Ueno}
\affiliation{Shibaura Institute of Technology, Toyosu campus, Koto-ku Tokyo, Japan}
\author{S. M. Yusuf}
\affiliation{Solid State Physics Division, Bhabha Atomic Research Centre, Mumbai-400085, India}
\affiliation{Homi Bhabha National Institute, Anushaktinagar, Mumbai-400094, India}
\author{R. S. Dhaka}
\email{rsdhaka@physics.iitd.ac.in}
\affiliation{Department of Physics, Indian Institute of Technology Delhi, Hauz Khas, New Delhi-110016, India}
\date{\today}                     

\begin{abstract}

We study the structural, magnetic, transport and electronic properties of LaCoO$_3$ with Sr/Nb co-substitution, i.e., La$_{(1-2x)}$Sr$_{2x}$Co$_{(1-x)}$Nb$_{x}$O$_3$ using x-ray and neutron diffraction, dc and ac-magnetization, neutron depolarization, dc-resistivity and photoemission measurements. The powder x-ray and neutron diffraction data were fitted well with the rhombohedral crystal symmetry (space group \textit{R$\bar{3}$c}) in Rietveld refinement analysis. The calculated effective magnetic moment ($\approx$3.85~$\mu_B$) and average spin ($\approx$1.5) of Co ions from the analysis of magnetic susceptibility data are consistent with 3+ state of Co ions in intermediate-spin (IS) and high-spin (HS) states in the ratio of $\approx$50:50, i.e., spin-state of Co$^{3+}$ is preserved at least up to $x=$ 0.1 sample. Interestingly, the magnetization values were significantly increased with respect to the $x=$ 0 sample, and the M-H curves show non-saturated behavior up to an applied maximum magnetic field of $\pm$70 kOe. The ac-susceptibility data show a shift in the freezing temperature with excitation frequency and the detailed analysis confirm the slower dynamics and a non-zero value of the Vogel-Fulcher temperature T$_0$, which suggests for the cluster spin glass. The unusual magnetic behavior indicates the presence of complex magnetic interactions at low temperatures. The dc-resistivity measurements show the insulating nature in all the samples. However, relatively large density of states $\approx$10$^{22}$ eV$^{-1}$cm$^{-3}$ and low activation energy $\approx$130~meV are found in $x=$ 0.05 sample. Using x-ray photoemission spectroscopy, we study the core-level spectra of La 3$d$, Co 2$p$, Sr 3$d$, and Nb 3$d$ to confirm the valence state. 
\end{abstract}
\maketitle

\section{\noindent ~Introduction}

The spin-state transition present in the LaCoO$_3$ is an open question for the researchers even after enormous theoretical as well as experimental efforts since many decades \cite{AbbatePRB93, KorotinPRB96, HaverkortPRL06, AltarawnehPRL12, AugustinskyPRL13, ChakrabartiPRM17,ShuklaPRB18}. In LaCoO$_3$, Co is in Co$^{3+}$ (3d$^6$) state and their ground state is characterized by the low spin electronic configuration [LS; S=0; t$^6_{2g}e^0_g$], whereas other excited states are defined as intermediate spin state (IS; S = 1; t$^5_{2g}e^1_g$) and high spin state (HS; S = 2; t$^4_{2g}e^2_g$). The ground state of LaCoO$_3$ gradually passes to IS, HS or mixed state of IS/HS with temperature, pressure, magnetic field, cationic substitution as well as with induced lattice strain \cite{IshikawaPRL04,EnglishPRB02,ZobelPRB02,AltarawnehPRL12,YamaguchiPRB97,SaitohPRB97,KozlenkoPRB07,VasquezPRB96, KrienerPRB04, SatoJPSJ14,StreltsovJPCM16,KarpinskyJPCM05,SerranoJPDA08,HammerPRB04, BaierPRB05,KrienerPRB09,YoshiiPRB03}. The energy difference between LS and IS state of the Co$^{3+}$ ion is rather small ($\sim$80~K in temperature scale), usually this energy difference increases with the reduction of CoO$_6$ octahedron volume and vice versa \cite{TroyanchukPSS05}, however above 80~K, Co ion is found in IS configuration showing Jahn-Teller (JT) distortion \cite{KorotinPRB96}. The Co$^{3+}$ possess larger ionic radius in IS state (0.56~\AA) than in the LS state (0.545~\AA) \cite{RadaelliPRB02,ShannonACA76,ShannonACB69}; therefore, an increase in unit cell volume favors transition from LS to IS state. The second transition at 500~K is attributed to an insulator to metal transition. The band structure calculations within the LDA+U approximation demonstrated that the IS state is lowest in energy after the first transition  \cite{KorotinPRB96,RaccahJAP68}. In contrast to the expectation from the simple ionic model, IS is stabilized by a strong p-d hybridization and possible orbital ordering in the e$_g$ shell of Co$^{3+}$ ions \cite{KorotinPRB96}. The redistribution of electrons between the t$_{2g}$ and e$_g$ levels results from a competition between the crystal field splitting energy ($\Delta_{CF}$) and the intra-atomic Hund's exchange energy (J$_{ex}$); however, both terms have comparable values in cobaltites. The crystal field splitting energy strongly depends on the Co-O bond length, therefore in a pure ionic picture the ground state is stabilized with $\Delta$ $>$ 3J$_{ex}$ and for HS state $\Delta$ $<$ 3J$_{ex}$. The competition between $\Delta$ and J$_{ex}$ can easily be controlled by cationic substitution either at La or Co site \cite{IshikawaPRL04,KrienerPRB04}. This has been observed in the behavior of LaCo$_{1-x}$Rh$_x$O$_3$ and La$_{1-x}$Sr$_x$CoO$_3$, in which the spin state transition disappears for certain compositions and Co ions remain/retain magnetic down to the lowest measured temperature \cite{KyomenPRB03, WuPRB03}.

The Sr$^{2+}$ ionic radius (1.44~\AA) is significantly greater than that of the La$^{3+}$ ion (1.36~\AA), so it is possible to anticipate that the stabilization of IS state of Co ions by substituting Sr$^{2+}$ ions at La$^{3+}$ site \cite{RaviJALCOM18}. We believe that Sr ions, whose radius is larger than that of La ions, increase the average Co-O distance in part of the CoO$_6$ octahedra, which favors transition of the Co$^{3+}$ ions from the LS to IS state. However, at such heterovalent substitution Co$^{4+}$ ions appear, leading to the ferromagnetic metallic ground state \cite{WuPRB03, SenarisJSSC95, ItohJMMM95}. The majority of researchers intended that ferromagnetism in cobaltites attribute to the phenomenon of double exchange interaction (between Co$^{3+}$--Co$^{4+}$), as comprehended in manganites \cite{ZenerPR151,ZenerPR251}. Other interesting case is that the hole doping (Ca, Sr, Ba at La site) in LaCoO$_3$ drives the system in metallic regime whereas the substitution of Nb at Co site impels towards the insulating regime \cite{ShuklaPRB18}. Therefore, the co-substitution of both cations on their respective sites in LaCoO$_3$ would alter its transport behavior. It is well reported that regardless of having larger ionic radii, Sr$^{2+}$ does not induce any structural transformation in the system up to 50\% concentration \cite{NamPRB99}. Whereas, our recent study reveals that the Nb$^{5+}$ substitution at Co site induces structural transformation of lower symmetry from the rhombohedral to orthorhombic, and then monoclinic at higher concentration \cite{ShuklaPRB18}. 

Here, we have simultaneously substituted Sr and Nb in place of La and Co in a ratio of 2:1, respectively. The substitution of Sr$^{2+}$ at La$^{3+}$ site changes the Co ion from Co$^{3+}$ to Co$^{4+}$, and the substitution of one Nb$^{5+}$ at Co site converts two Co$^{3+}$ into the Co$^{2+}$ ions \cite{ShuklaPRB18, WuPRB03}. Therefore with co-substitution of Sr and Nb in 2:1 ratio, the valence state of Co ion can be maintained in 3+. In order to understand the effect of co-substitution in La$_{(1-2x)}$Sr$_{2x}$Co$_{(1-x)}$Nb$_{x}$O$_3$, we study the structural, magnetic, transport and electronic properties. The Rietveld refinement of x-ray and neutron diffraction data reveal the rhombohedral symmetry for all the samples and an increase in the unit cell volume with higher substitution. The derived values of effective magnetic moment confirm the 3+ valence state of Co ions, which is consistent as co-substitution is expected to maintain Co ions in 3+ state. We further confirm the valence state of Co ion using photoemission spectroscopy. The extracted characteristic frequency from ac-susceptibility data exhibits slower spin dynamics of the system and suggests that the origin of spin glass state is not from individual spins rather related to their clusters. The transport properties manifest that co-substitution drives system towards the insulating regime; fitted data for the possible conduction mechanisms provide the values of activation energy as well as density of states near the Fermi level. 

\section{\noindent ~Experimental}

Polycrystalline samples of La$_{(1-2x)}$Sr$_{2x}$Co$_{(1-x)}$Nb$_{x}$O$_3$ were synthesized by the conventional solid state solution method. We used strontium carbonate (SrCO$_3$), cobalt oxide (Co$_3$O$_4$) and niobium oxide (Nb$_2$O$_5$) as purchased, whereas purchased lanthanum oxide (La$_2$O$_3$) powder was dried at 700 $^0$C for 12 hrs prior to use. Stoichiometric amount of starting materials (all $\ge$99.9\% from Sigma/Alfa) were thoroughly mixed with the help of mortar-pestle and reacted at 1100$^0$C for 12 hrs in air, pellets of calcined powder were cold pressed at 2000 psi and sintered in air at 1475$^0$C for 6 -- 10 hrs \cite{ShuklaPRB18,OygardenJSSC12}. 

X-ray diffraction data were collected with a CuK$\alpha$ radiation ($\lambda$ = 1.5406 \AA ) using Panalytical x-ray diffractometer, magnetic and transport measurements were done with physical property measurement system (PPMS) from Quantum Design, USA. Ac-susceptibility measurements were carried out in a CRYOGENIC make, liquid He based PPMS at BARC, India. We analyzed the XRD data by Rietveld refinement using FullProf package and the background was fitted using the linear interpolation between the data points. We recorded the powder neutron diffraction data at temperatures 3 and 300~K for bulk La$_{(1-2x)}$Sr$_{2x}$Co$_{(1-x)}$Nb$_{x}$O$_3$ ($x=$ 0.025, 0.05, and 0.15) samples, using the powder diffractometer PD-I ($\lambda$ = 1.094~\AA~) at the Dhruva reactor, Trombay, Mumbai, India. One dimensional neutron depolarization measurements over the temperature range of 4--300 K, to identify the  ferromagnetic clusters, were carried out using the polarized neutron spectrometer (PNS) at the Dhruva reactor, under an applied magnetic guide field of 50~Oe. For the neutron depolarization measurements, polarized neutrons ($\lambda$ = 1.205~\AA~) were produced and analyzed by using magnetized Cu$_2$MnAl (111) and Co$_{0.92}$Fe$_{0.08}$ (200) single crystals, respectively. The two different states (up and down) of the incident neutron beam polarization were achieved by a $\pi$ flipper just before the sample. The polarization of the neutron beam was determined by measuring the intensities of neutrons in non-spin flip and spin flip channels with the flipper off and on (flipping ratio, R), respectively. The core-level spectra of La 3$d$, Co 2$p$, Sr 3$d$ and Nb 3$d$ were recorded at room temperature using a monochromatic Al-K$\alpha$ source (energy resolution = 0.5~eV), in a base pressure of 5 $\times$ 10$^{-10}$ mbar. We used a charge neutralizer to compensate the charging effect in these insulating samples.

\section{\noindent ~Results and Discussion}

\subsection{\noindent ~A. Room temperature x-ray and neutron diffraction}

\begin{figure}
\includegraphics[width=3.35in]{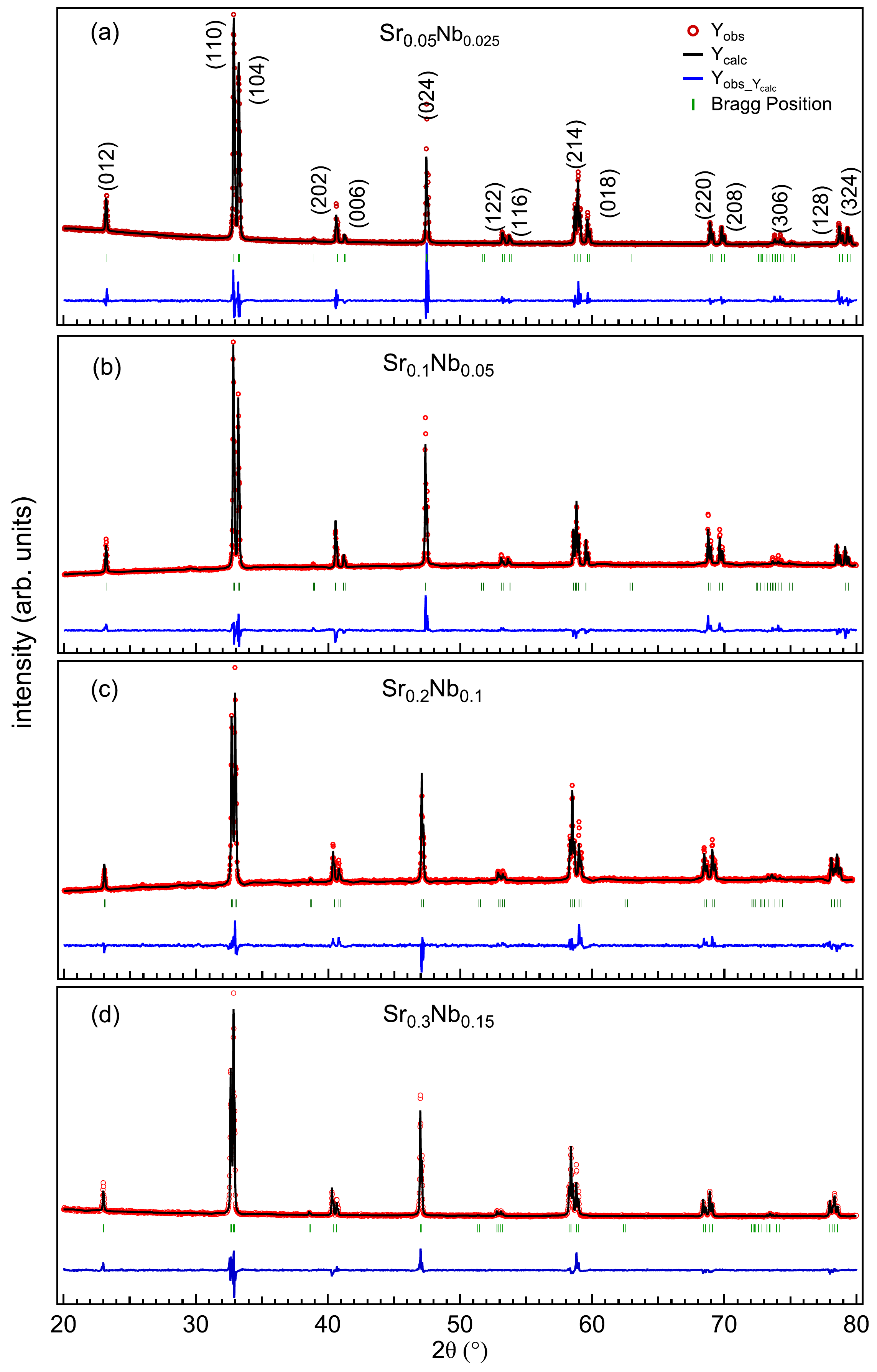}
\renewcommand{\figurename}{Figure}
\caption{Rietveld refined powder x-ray diffraction data of La$_{(1-2x)}$Sr$_{2x}$Co$_{(1-x)}$Nb$_{x}$O$_3$ samples, (a) $x$ = 0.025, (b) $x$ = 0.05, (c) $x$ = 0.1, and (d) $x$ = 0.15. The Rietveld refinement profiles, the fitted Bragg peak positions, and the residual are shown in black solid lines, short vertical bars and blue lines in the bottom of each panel, respectively.}
\label{fig:XRD}
\end{figure}

The crystal structure of La$_{(1-2x)}$Sr$_{2x}$Co$_{(1-x)}$Nb$_{x}$O$_3$ samples has been investigated by the Rietveld refinement of both x-ray and neutron diffraction data measured at room temperature. The lattice parameters were determined from the refinement of x-ray diffraction (XRD) patterns and were used as a starting point for the refinement of neutron diffraction (ND) data. Due to sensitivity of neutron towards lighter atom, the positions of oxygen were obtained from the refinement of neutron diffraction data and were kept fix during the final stage refinement of x-ray diffraction data. The Rietveld refined x-ray and neutron diffraction patterns are presented in Figures~1 and 2, respectively. The refinement confirms single phase formation of these samples in rhombohedral structure (space group \textit{R$\bar{3}$c} \#167). The Rietveld refined lattice parameters and unit cell volume show systematic increment with increasing the amount of substitution, as presented in table~I. Also, in table~I, we present pseudocubic (pc) lattice parameters calculated using a$_{pc}$ = a/$\sqrt{2}$, b$_{pc}$ = b/$\sqrt{2}$, c$_{pc}$ = c/$\sqrt{12}$, and unit cell volume (V$_{pc}$) for La$_{(1-2x)}$Sr$_{2x}$Co$_{(1-x)}$Nb$_{x}$O$_3$. Recently, it has been reported that the Nb substitution at Co site in LaCoO$_3$ induces the structural transition \cite{OygardenJSSC12,ShuklaPRB18}, which can be attributed to the difference in the ionic radius of Nb$^{5+}$ and Co$^{3+}$ ions. In LaCoO$_3$, Co$^{3+}$ ions have ionic radii of 0.545~\AA~in LS state and 0.61~\AA~ in HS state, but Nb$^{5+}$ ions have ionic radii of 0.64~\AA~\cite{ShannonACB69, ShannonACA76}. Due to the larger ionic radii of Nb ions, volume of octahedra around Co site increases and results in the modification of crystal structure. On the other hand, larger Sr$^{2+}$ substitution at La site in LaCoO$_3$ will not change the structure up to $x=$ 0.5 \cite{NamPRB99}. Therefore, the co-substitution of Sr$^{2+}$ and Nb$^{5+}$ in LaCoO$_3$ may generate the possibility of structural modification attributed to the perturbation of volume of polyhydra around both the sites. However, we observe that with co-substitution up to $x=$ 0.15, there is no structural transition and we were able to fit the XRD and ND data with the rhombohedral space group \textit{R$\bar{3}$c} only, as clearly visible in Figures~1 and 2. 

\begin{table*}
		\centering
		\label{tab:Rietveld}
		\renewcommand{\tablename}{Table}
		\renewcommand{\thetable}{\arabic{table}}
		\caption{The Rietveld refined parameters of La$_{(1-2x)}$Sr$_{2x}$Co$_{(1-x)}$Nb$_{x}$O$_3$ samples; the lattice parameters and unit cell volume obtained from XRD analysis; the bond-lengths and bond-angles obtained from neutron diffraction, measured at 300~K.}
		\vskip 0.2cm
		\begin{tabular}{|c|c|c|c|c|c|c|ccc|}
		\hline
		XRD&$\chi^2$&R$_p$&R$_{wp}$&a=b(\AA) &c (\AA) &V(\AA$^3$) & a$_{pc}$=b$_{pc}$ &c$_{pc}$ & V$_{pc}$\\
				 $x$(\%)& &(\%) & (\%) &&&&(\normalfont\AA) &(\normalfont\AA) &(\normalfont\AA$^3$)\\
		\hline
		0&1.75&3.31&4.57&5.4416(13)&13.0900(2)&335.679(9)&3.848&3.778&55.95\\
2.5&4.19&4.16&6.85&5.4449(13)&13.1195(4)&336.841(15)&3.850&3.783&56.14\\
5&3.2&3.52&5.35&5.4554(13)&13.1448(4)&338.801(15)&3.858&3.805&56.64 \\
10&2.3&2.91&4.33&5.4766(16)&13.2594(5)&344.407(19) & 3.871& 3.841&57.55 \\
15&2.8&4.21&6.23&5.4816(12)&13.2996(3)&346.087(14)& 3.899& 3.869&58.82 \\
		\hline
		\end{tabular}

		\centering

	\begin{tabular}{|c|c|c|c|c|ccccccc|}
		\hline
		neutron&$\chi^2$&R$_p$&R$_{wp}$& Co-O (\normalfont\AA)&Co-O-Co&O-Co-O&La-O (\normalfont\AA)&La-O (\normalfont\AA)&La-O (\normalfont\AA)&La-Co (\normalfont\AA)& La-Co (\normalfont\AA)\\
				$x$ (\%)&&(\%)&(\%)&$\times$6&&&$\times$3&$\times$3&$\times$6&$\times$2&$\times$6\\
		\hline
2.5&2.27&1.42&2.0&1.9439(2)&163.029(1)&91.225(5)&3.0204(2)&2.4467(2)&2.7196(2)&3.2924(4)&3.3420(2)\\
5&4.85&1.84&2.77&1.9440(1)&164.465(3)&91.052(6)&2.9996(2)&2.4741(2)&2.7239(3)&3.3047(4)&3.3467(2)\\
15&2.28&1.52&2.03&1.9554(1)&165.693(1)&90.817(6)&2.9963(2)&2.5093(2)&2.7455(2)&3.3381(4)&3.3678(2)\\
		\hline
		\end{tabular}
\end{table*}

\begin{figure}[h]
\includegraphics[width=3.3in]{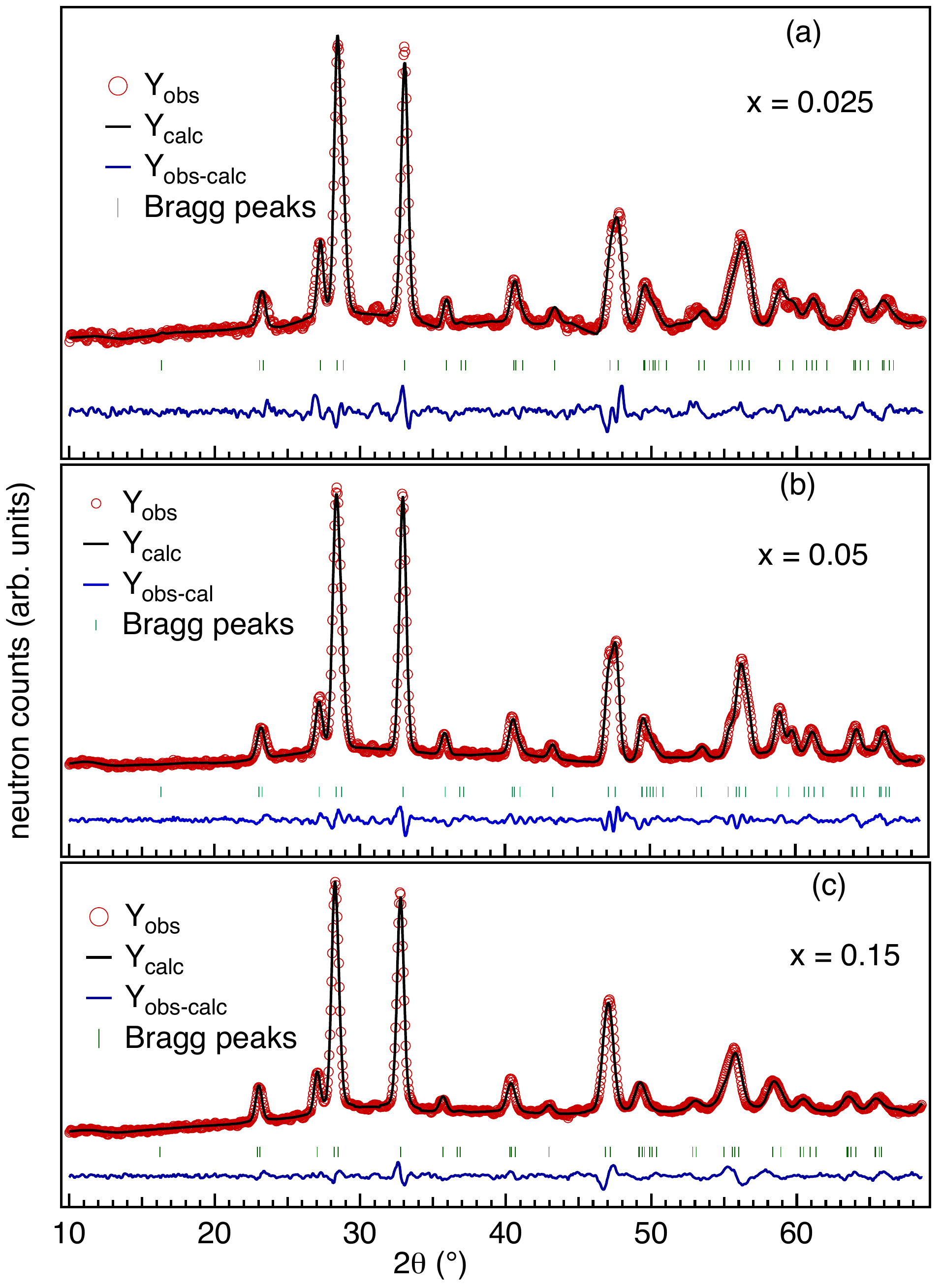}
\renewcommand{\figurename}{Figure}
\caption{Rietveld refined neutron powder diffraction data for the  La$_{(1-2x)}$Sr$_{2x}$Co$_{(1-x)}$Nb$_{x}$O$_3$ samples, (a) $x$ = 0.025, (b) $x$ = 0.05, and (c) $x$ = 0.15, measured at room temperature. The Rietveld refinement profiles, the fitted Bragg peak positions, and the residual are shown in black solid lines, short vertical bars and blue lines in the bottom of each panel, respectively.}
\label{fig:NPD}
\end{figure}

In Figures~\ref{fig:NPD}(a--c) we show the Rietveld refined powder neutron diffraction patterns measured at 300~K for the $x=$ 0.025, 0.05, and 0.15 samples, respectively. All the observed patterns were fitted by considering only the nuclear phase. In ABO$_3$ type perovskites the A site has cuboctahedral symmetry and here it has three groups of La--O bond-lengths, namely small (3 fold degenerate), intermediate (6 fold degenerate) and long (3 fold degenerate). The Rietveld refined parameters like bond-lengths and bond-angles from the Neutron diffraction for La$_{(1-2x)}$Sr$_{2x}$Co$_{(1-x)}$Nb$_{x}$O$_3$ are summarized in table~I. 
We found that the long bond decreases, while the other two groups of bonds increases with the substitution. This indicates that the long-bond compensate the effect of substitution of larger cations to preserve the crystal structure similar to the parent ($x=$ 0) sample. The B site has an octahedral symmetry and sixfold degenerate Co--O bonds. Here the Co--O bond length increases and slightly larger than LaCoO$_3$ (1.934~\AA) at 300~K \cite{RadaelliPRB02}, while the octahedra changes its bond-angles i.e., O--Co--O decreases with the substitution.

Octahedral distortions are reflected in the tilting of BO$_6$ from the vertical axis and defined as $\phi_1$ = (180-$\Theta$)/2, where $\Theta$ is defined as B-O-B bond angle along the z-axis. The tilting of octahedra is also quantified through the pseudo-cubic parameters c$_{pc}$ and $a_{pc}$, defined as c/$\sqrt{12}$ and $a$/$\sqrt{2}$, respectively. The distortion parameter $\Delta_d$ is defined as (1/12)$\Sigma_{n=1-12}$ [($d_n-d_{avg})/d_{avg}]^2$, where d$_{avg}$ is average bond length of La--O. The calculated values of $\Delta_d$ for the cuboctahedra are presented in table~\ref{tab:Strain}, where we can see that the tetragonal strain decreases with increasing the substitution. In order to obtain the information about the oxidation state of metal ion, we have calculated bond valence sum (BVS) from the bond-lengths of the octahedra without any assumptions. The BVS is generally calculated using an empirical relation based on the Pauling's ``2$^{nd}$ Rule" of electrostatic valence, which postulates that the total strength of the valence bond, which reaches an ion from all the surrounding atoms is equal to the charge of the ion \cite{PaulingJACS29,OgielskiPRB8587}. $$V_j = \sum_{j} s_{ij}$$ where the sum is over neighboring $j$ atoms with each bond between atoms $i$ and $j$ having bond valence s$_{ij}$. Brown and Altermatt suggested the exponential form in 1985 \cite{BrownAC85} $s_{ij} = [exp(R_0-R_{ij})/b]$, where R$_0$ is termed as unit valence bond-length, and therefore a unique value for each metal-ion pair was tabulated. Here, R$_{ij}$ is the experimentally observed bond-length between the $ij$ metal-ion pair and $b$ is an empirical constant, generally taken as 0.37 \cite{BrownAC73}. The bond valence parameter (R$_0$) depends on the metal-ion pair, valency of metal ion and the coordination number of the metal ion. The calculated BVS values from room temperature neutron data are presented in the table \ref{tab:Strain} for $x=$ 0.025, 0.05 and 0.15 samples. These calculated BVS values signifies the average effect of the charge with the substitution around the A-O$_{12}$ and B-O$_6$ polyhydra. Since, the substitution of Sr and Nb changes the average valency on the A ($v_{\rm A}$) and B ($v_{\rm B}$) sites, but to maintain the charge neutrality a charge transfer occurs between the A and B sites. The charge deficit for the each sample is calculated with the formula $\delta$ = 3--($\langle$v$_{\rm A}\rangle$+$\langle$v$_{\rm B}\rangle$)/2, which directly correlates with the valence charge and termed as the non-stoichiometry. 
\begin{table}[h]
  \centering
  \renewcommand{\tablename}{Table}
		\renewcommand{\thetable}{\arabic{table}}
  \caption{The strain parameter is e$_t$ = (c$_{pc}$-a$_{pc}$)/(c$_{pc}$+a$_{pc}$). c$_{pc}$ and a$_{pc}$ are the pseudocubic lattice constants given by c/$\sqrt{12}$ and a/$\sqrt{2}$. The distortion parameter $\Delta_d$ for the La-O polyhydra. The $\phi_1$ is the tilting angle about the z-axis for BO$_6$ octahedra. The average bond valence sum (BVS) for the B and A sites are v$_{\rm B}$ and v$_{\rm A}$, respectively.}
  \label{tab:Strain} 
  \vskip 0.2 cm
   \begin{tabular}{|c|c|c|c|c|c|c|c|}
  	\hline
   $x$ (\%)& $c_{pc}$/a$_{pc}$ & e$_t$ &$\phi_1$ & $\Delta_d$ &v$_{\rm B}$&v$_{\rm A}$\\
   &  & $\times$10$^{-3}$& (deg) &  $\times 10^{-2}$&&\\
    \hline
    2.5& 0.98&-8 &8.48 & 6.6&3.103&3.096\\
    \hline
    5 & 0.985 &-7 & 7.7&  5.5&3.103&2.996\\
    \hline
    15 & 0.99&-4.5 & 7.15&  4.7&3.008&2.802\\
    \hline
  \end{tabular}
\end{table}
For $\delta \neq $0, the valence charge at oxygen site would be 6$\pm \delta$ in the unit cell. We have calculated the saturated bond lengths, below this coordination polyhydra cannot occupy more charge. The saturated bond valence ($\nu_s$) can be calculated using $\nu_s$ = e$^{(R_0-R_s)/b}$, where R$_s$ is the saturated bond valence parameter. The difference between the measured and saturated valence can be calculated, which gives the total valence deviation $\Delta\nu$=$\Delta\nu_{A-S}$--$\Delta\nu_{B-S}$ and demonstrates the balancing of the valence charge between the B-O$_6$ and A-O$_{12}$ polyhydra and vice-versa. We have plotted $\Delta\nu$ with the substitution concentration (not shown) and a linear fit of this curve crosses the $y=$ 0 axis near the $x=$ 0.12 value (i.e. around $x$ equal to 12\% concentration). This means that for $x >$ 0.12, coordination polyhydra of A and B declines equally from the saturation. The estimated value of $\delta$ is found to be positive (about 0.09) only for the $x=$ 0.15 sample. On the other hand, Brown and Shannon tested and reported that the bond valence method has accuracy around 5-7\% for ionic compounds \cite{BrownAC73}. These obtained parameters from the structural analysis (particularly Co--O length and larger Co--O--Co angle) influence the magnetic properties of these samples \cite{DurandJPCM13, DurandJPCM15}.

\subsection{\noindent ~B. Magnetization and magnetic susceptibility}

\begin{figure}[h]
\centering
\includegraphics[width=3.5in]{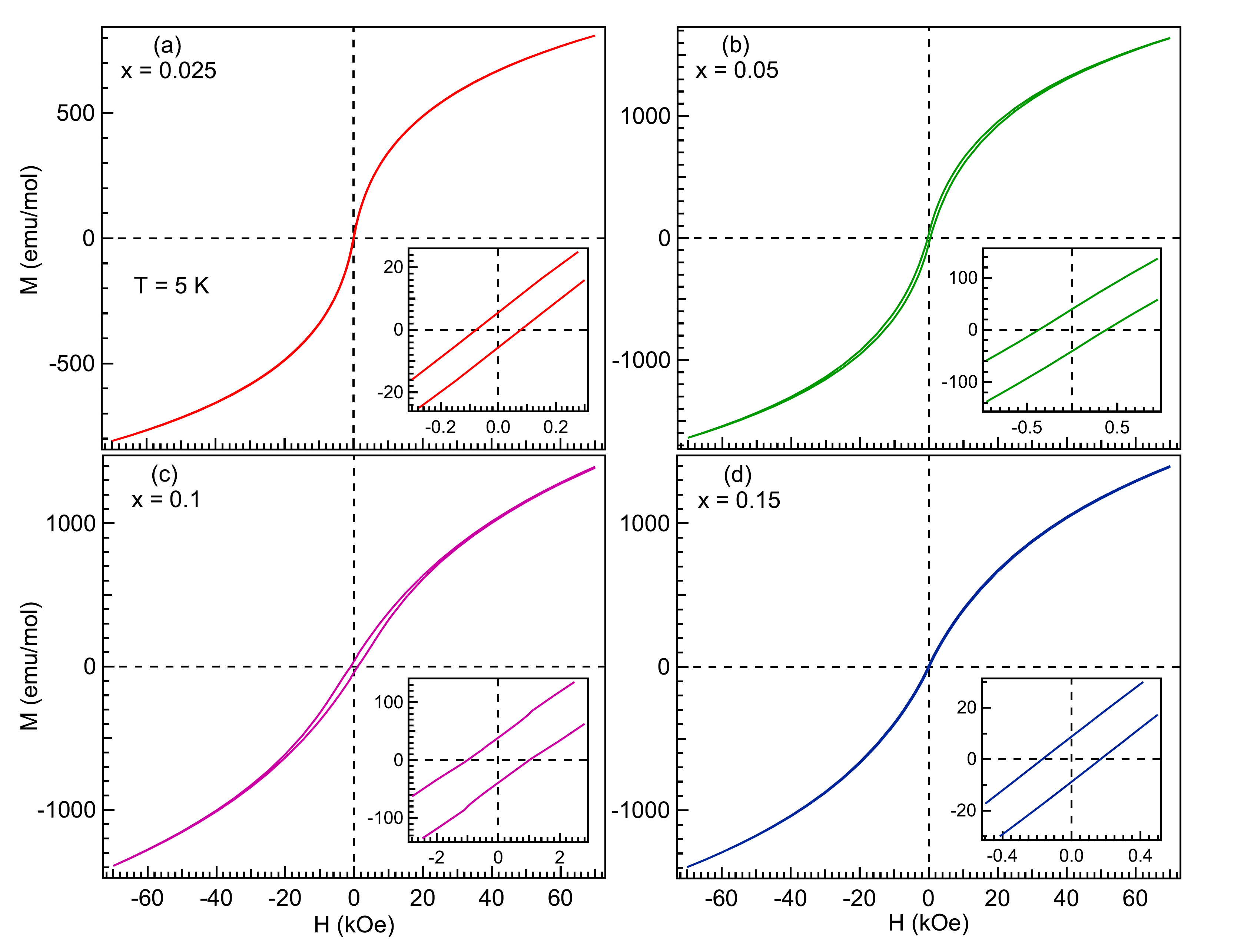}
\renewcommand{\figurename}{Figure}
\caption{Isothermal magnetization (M--H) data for the La$_{(1-2x)}$Sr$_{2x}$Co$_{(1-x)}$Nb$_{x}$O$_3$ samples, (a) $x$ = 0.025, (b) $x$ = 0.05, (c) $x$ = 0.1, and (d) $x$ = 0.15, measured at 5~K, inset in each panel shows the magnetic behavior near origin.}
\label{fig:MH}
\end{figure}

In order to study the magnetic behavior, we have performed isothermal magnetization (M--H) measurements at 5~K with variation of applied magnetic field upto $\pm$70~kOe, the M-H loops for all the samples are presented in Figures~\ref{fig:MH} (a--d) where the inset in each panel shows the zoomed view of M--H close to zero field. 
Recently, we observed the maximum value of magnetization $\sim$170~emu/mol for the parent ($x=$ 0) sample at highest applied magnetic field with the coercivity value of 1.35~kOe \cite{ShuklaPRB18}. On the other hand, here as we increase the concentration of co-substitution gradually, the maximum magnetization increased to $\sim$800 emu/mol (for the $x=$ 0.025), but the hysteresis is almost disappeared; [see Figure~\ref{fig:MH}(a)]. Further, for the $x=$ 0.05 sample the magnetization increases to $\sim$1600~emu/mol, which is around ten times higher than that of the $x=$ 0 sample. For the $x=$ 0.1 sample this value found to be $\sim$1400~emu/mol and concurrently an increase in the hysteresis. Further, for the $x=$ 0.15 sample there is no change in the magnetization value ($\sim$1400~emu/mol), but the hysteresis found to be negligible. It is interesting to note that there is a large increase in the magnetization with co-substitution of Sr and Nb in 2:1 ratio, where we expect the Co ion maintains 3+ valence state like in the parent ($x=$ 0) sample. In LaCoO$_3$, an increase in the average Co--O bond length of the CoO$_6$ octahedra promote a spin-state transition from LS to IS state \cite{WuPRB03}. In the present case this behavior is evident as Co-O distance increases with the co-substitution in comparison to the $x=$ 0 sample and this favors the possibility of similar spin-state transition.  Although the nearest Co ions in IS state can interact ferromagnetically through the e$^1$-O-e$^0$ superexchange, as described in ref. \cite{YanPRB04}; however, the substitution of Nb ions, which act as the magnetic dilution in the system, decreases the probability of long-range magnetic ordering in the system. Also, only about 50\% Co ions are in IS state, discussed later. The cumulative effect of these interactions reflects in the M--H curves, where first coercivity increases up to $x=$ 0.1 sample and then decreases for the $x=$ 0.15 sample. This is consistent as this Nb concentration is found to be close to the percolation limit, also found in the BVS results, for the magnetic dilution.

\begin{figure}[h]
\centering
\includegraphics[width=3.5in]{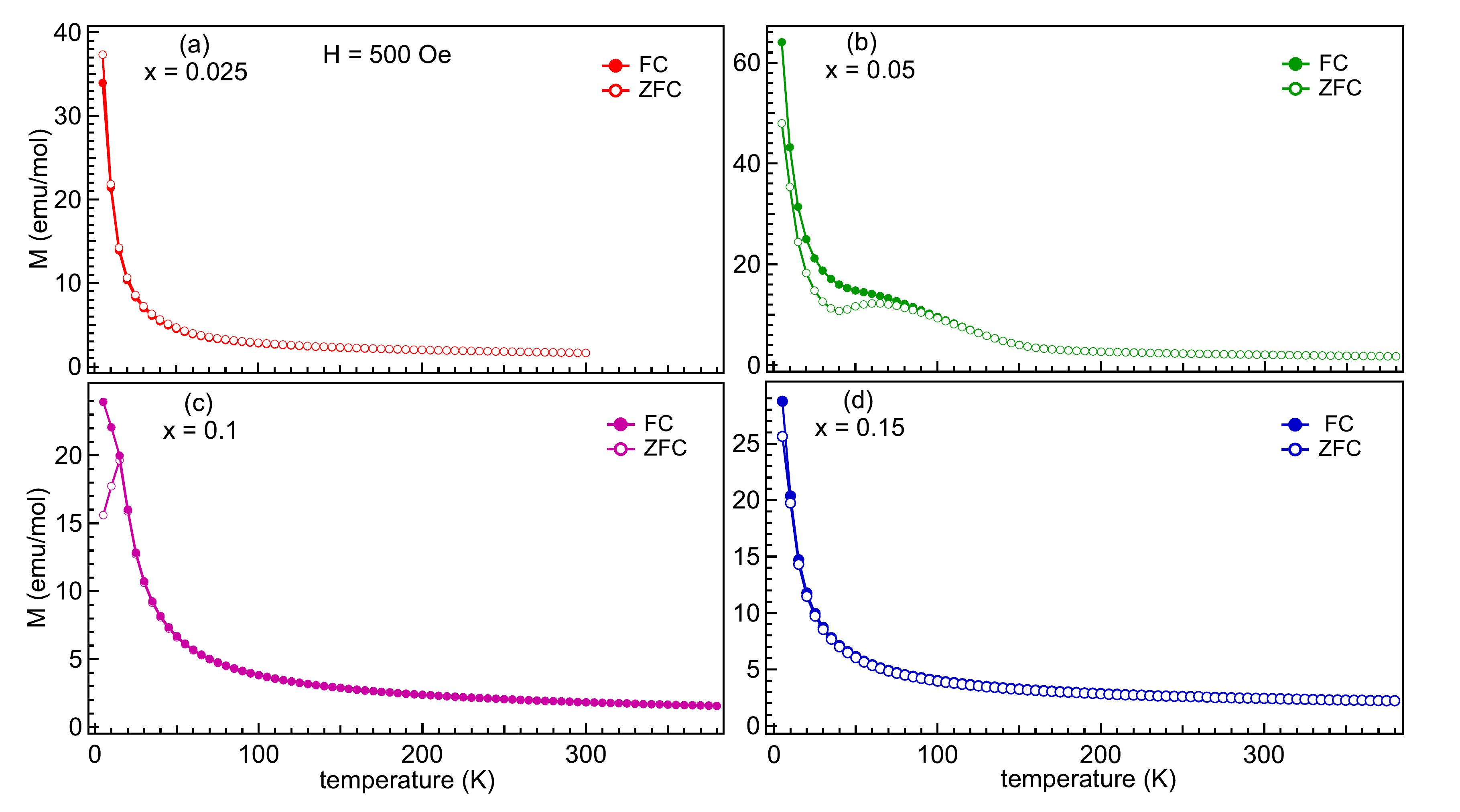}
\renewcommand{\figurename}{Figure}
\caption{Temperature dependent ZFC--FC curves for the La$_{(1-2x)}$Sr$_{2x}$Co$_{(1-x)}$Nb$_{x}$O$_3$ samples, (a) $x$ = 0.025, (b) $x$ = 0.05, (c) $x$ = 0.1, and (d) $x$ = 0.15, measured at 500~Oe field.}
\label{fig:MT}
\end{figure}

\begin{figure}
\centering
\includegraphics[width=3.5in]{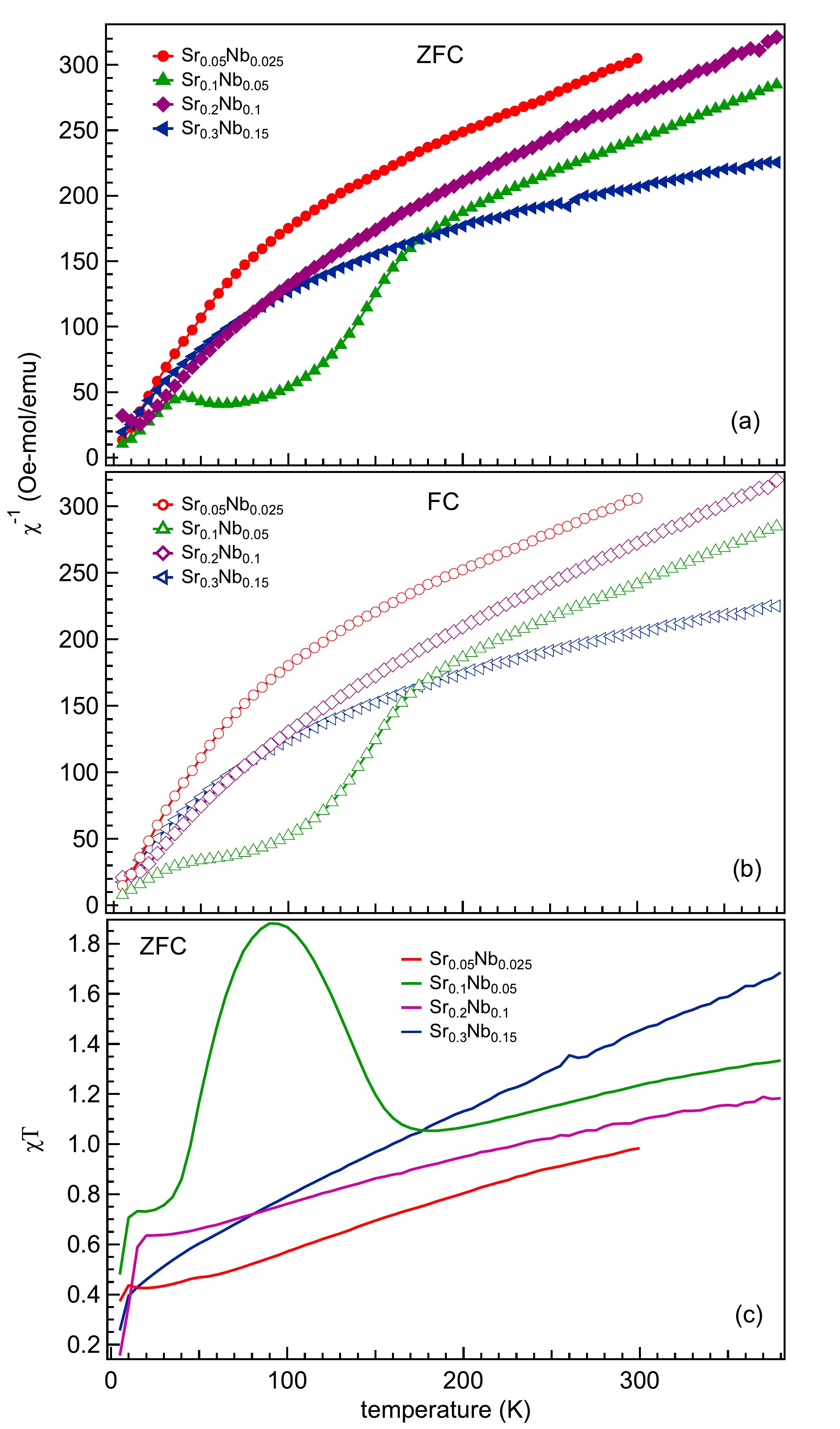}
\renewcommand{\figurename}{Figure}
\caption{The $\chi^{-1}$ plots for the La$_{(1-2x)}$Sr$_{2x}$Co$_{(1-x)}$Nb$_{x}$O$_3$ samples as a function of temperature in (a) ZFC and (b) FC modes, and (c) the plot between $\chi$T (ZFC) vs temperature for all the samples.}
\label{fig:CW}
\end{figure}

For further understanding the magnetic susceptibility and possible spin states of Co ion, we have studied the temperature dependence of the magnetization M(T) by performing the measurements in the zero-field cooled (ZFC) and field cooled (FC) modes in the temperature range of 5--380~K [see Figures~\ref{fig:MT}(a--d)]. 
We observe that there is no bifurcation between ZFC--FC curves, but a sharp increase in the magnetization below $\sim$10~K temperature is clearly present for the $x=$ 0.025, 0.15 samples. This behavior suggest the presence of Curie-Weiss paramagnetic behavior and/or surface related magnetism, which dominates in the low temperature regime and suggest the absence of long-range ordering in these samples \cite{YanPRB04}. On the other hand, the M(T) data for the $x=$ 0.05 sample exhibit a bifurcation between ZFC-FC curves below 80~K, and a cusp in the ZFC curve near 40~K and then the magnetization increases sharply to $\sim$60~emu/mol. For the $x=$ 0.1 sample, we observe a bifurcation below 15~K and the ZFC magnetization decreases while FC increases at lower temperatures. The observed unconventional behavior in the magnetization for the $x=$ 0.05 sample could be due to the presence of temperature induced spin-state transition, as observed for the parent sample LaCoO$_3$ \cite{AndroulakisPRB01}. The bifurcation between the FC and ZFC magnetization curves for the $x=$ 0.05 and 0.1 samples indicate the possible presence of spin-glass phase \cite{GuoPRB16}, which further investigated using ac-magnetic susceptibility measurements (discussed later). We have analyzed the $\chi^{-1}$ vs. T curves of all the samples in the range of 220 to 380~K [see Figures~\ref{fig:CW} (a, b)] by the Curie-Weiss law using the following equation $$\chi^{-1} = \frac{3k_B}{\mu_{\rm eff}^2} (T-\theta_{\rm CW})$$ and evaluated the effective magnetic moment ($\mu_{\rm eff}$) and the Curie-Weiss temperature ($\theta_{\rm CW}$) \cite{ShuklaPRB18}. 

Within the approximation of only spin magnetic moment, average spin (S$_{\rm av}$) is calculated from the experimentally observed values with the formula $ \mu_{\rm eff} = 2\sqrt{S(S+1)}$, which was resultant of different spin state contributions of Co ions. The values obtained from the analysis are presented in Table \ref{tab:MH} with the different combination of spin states of Co, responsible for the magnetization present in the system. Note that the spin state of Co$^{3+}$ can be preserved in intermediate and high spin-states in the 50:50 ratio, as was reported for the parent sample \cite{ShuklaPRB18,TroyanchukaPSS06}. 
\begin{table}[h]
  \centering
  \renewcommand{\tablename}{Table}
		\renewcommand{\thetable}{\arabic{table}}
  \caption{Experimentally determined fitting parameters, Curie constant (C) in emu K mol$^{-1}$, effective magnetic moment ($\mu_B$) and calculated average spin (S$_{av}$) for La$_{(1-2x)}$Sr$_{2x}$Co$_{(1-x)}$Nb$_{x}$O$_3$ samples.}
  \label{tab:MH}
  \vskip 0.2 cm
   \begin{tabular}{|c|c|c|c|c|}
  	\hline
   $x$ & $\theta_{C_W} (K)$ & C &$\mu_{\rm eff}$ (exp) & S$_{\rm av}$ (exp)\\
    \hline
0 &-220 &1.85 &3.85 & 1.50 \\
    \hline
    0.025&-264 &1.84 &3.84 & 1.49 \\
    \hline
    0.05 &-157 &1.89  &3.88 & 1.51 \\
    \hline
    0.1&-155  &1.68 &3.66 & 1.4 \\
    \hline
    0.15 &-465 &3.73 &5.46 & 2.3\\
    \hline
  \end{tabular}
\end{table}
Thus co-substitution of Sr and Nb in the LaCoO$_3$ preserves the valence states and spin-states at least up to the $x=$ 0.1 sample. It is consistent as even small amount of Co$^{2+}$ enhance the $\mu_{eff}$ significantly from $\approx$ 3.85~$\mu_B$ to 4.65~$\mu_B$ \cite{ShuklaPRB18}, which is not the case here up to $x=$ 0.1 sample. On the other hand, it is interesting to note that the magnetization is significantly enhanced (see Figure~3) as compared to the parent sample \cite{ShuklaPRB18}. This found to be unconventional and justify the need to further investigate the local structure and magnetism using absorption spectroscopy and magnetic circular dichroism studies at synchrotron radiation facility. However, the observed large value of $\mu_{eff}$ for the $x=$ 0.15 sample can be due to the Nb concentration as it is near the percolation limit and can induce Co$^{2+}$ valence states. It is also found from the BVS calculations that $x\ge$0.12 will be overcharged. In the $\chi\cdot$T vs. temperature plots; see Figure~\ref{fig:CW}(c), we found that all samples manifest a linear behavior and then sharp decrease below 10~K, whereas for the $x=$ 0.05 sample a broad peak centred $\sim$80~K is observed. Interestingly, Androulakis {\it et al.} observed a very similar behavior in the parent LaCoO$_3$ sample with the application of magnetic field (0.1~Tesla) \cite{AndroulakisPRB01}. Interestingly, the authors found an evidence of ferromagnetic and antiferromagnetic interactions along with spin state transition at low temperatures \cite{AndroulakisPRB01}. The observed downturn in $\chi^{-1}$ vs. T curves [Figures~5(a, b)] at around ~80~K [a broad peak in Figure~5(c)] in the $x=$ 0.05 sample indicate a spin-state transition and presence of magnetic interactions at low temperatures \cite{AndroulakisPRB01}.

\begin{figure}
\centering
\includegraphics[width=3.5in]{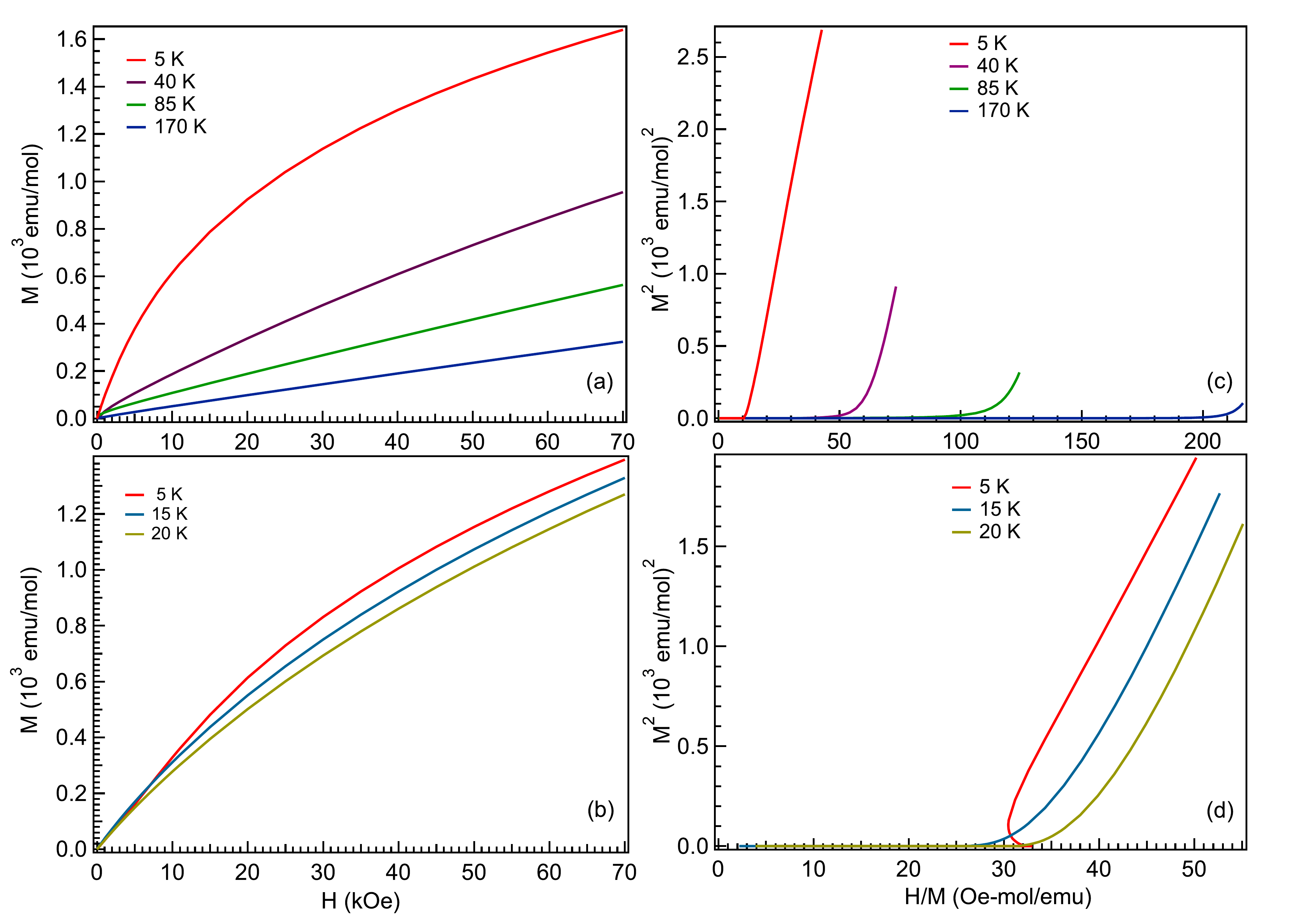}
\renewcommand{\figurename}{Figure}
\caption{(a, b) The virgin curves (M-H) and (c, d) Arrott plots (M$^2$ vs H/M) of (a, c) $x=$ 0.05, and (b, d) $x=$ 0.1 samples, at different temperatures across the transition.}
\label{fig:Arrott}
\end{figure}

Therefore, in order to probe the magnetic behavior across the bifurcation observed in the ZFC-FC data, we have measured the virgin curves (M vs. H) across the transition at various temperatures and plotted Arrott plots (M$^2$ vs. H/M) for $x=$ 0.05 and 0.1 samples, as shown in Figures~\ref{fig:Arrott}(a--d). From these M$^2$ vs. H/M curves, by fitting the high-field region with the straight line, which if have an intercept on the y-axis indicates a ferromagnetic interaction and a non-ferromagnetic interaction when an intercept on the x-axis. In these samples a linear fit gives an intercept on the x-axis for all samples which suggest a non-ferromagnetic interaction in the system, which is consistent with the M--H curves showing a non-saturating nature upto $\pm 70$~kOe. Recently, we have observed that in LaCoO$_3$, Nb substitution at Co site promotes short range ordering in the system for a large range of substitution ($x \leq$ 0.2) \cite{ShuklaPRB18,TroyanchukaPSS06}, and decreases the electrical conductivity as well. However, it has been reported that Sr substitution at La site induces the long range ordering in the system, an insulator to metal transition for $x=$ 0.18 and a metallic ground state for $x \geq$ 0.2 \cite{WuPRB03}. In the present case the Curie-Weiss temperatures are negative indicating antiferromagnetic interactions, whereas a small hysteresis in the isothermal magnetization curves indicate weak ferromagnetism along with the non-saturating behavior is a consequence of the antiferromagnetic interaction. This behavior suggests the coexistence of antiferromagnetic and ferromagnetic ordering and competition between them with co-substitution of Sr and Nb in LaCoO$_3$. This is consistent as the Co--O--Co bond angle ($\gamma$) values extracted from the neutron diffraction (see Table~I) are slightly larger than the critical value $\gamma_c$$\approx$163$\degree$ for LaCoO$_3$. The larger values of $\gamma$ in the present case is favorable for the presence of magnetism in these samples \cite{BelangerJPCM16, DurandJPCM13} probably due to internal strain as also the case in nanoparticles and thin films of LaCoO$_3$ \cite{DurandJPCM15, FuchsPRB08}. 

\subsection{\noindent ~C. ac-magnetic susceptibility}

\begin{figure}[h]
\centering
\includegraphics[width=3.4in]{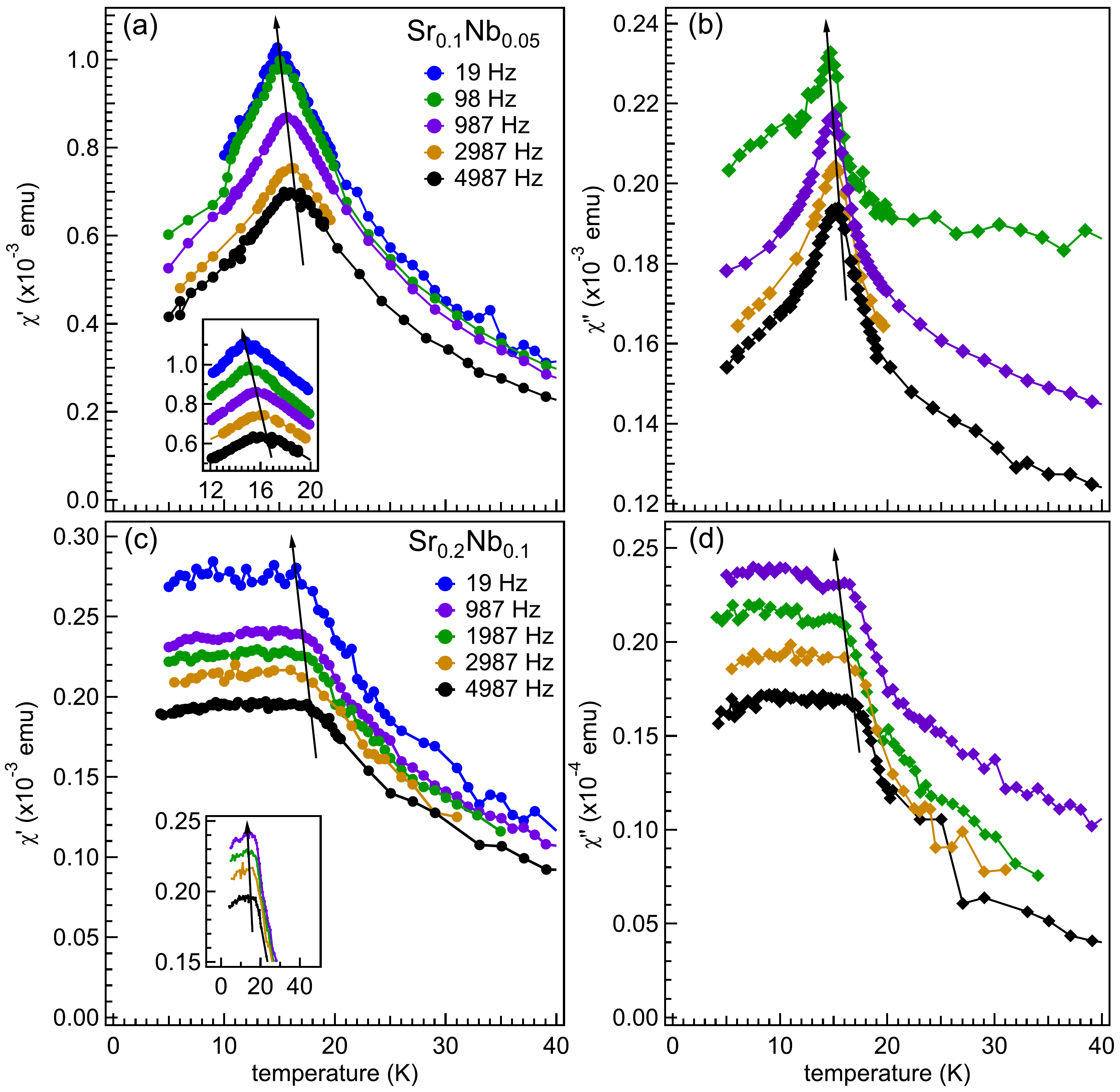}
\renewcommand{\figurename}{Figure}
\caption {Temperature dependent ac-susceptibility measurements (a, c) real part, and (b, d) imaginary parts of the $x=$ 0.05 and 0.1 samples, respectively; inset in (a) shows the shift in the peak position with the change in the frequency and inset in (c) shows that the susceptibility decreases after the freezing temperature T$_f$.}
\label{fig:ACS}
\end{figure}

Further we performed ac-susceptibility measurements for the $x=$ 0.05 and 0.1 samples. In Figures~\ref{fig:ACS}(a, b), we present the ac-susceptibility data for the $x=$ 0.05 sample, which show peaks centered at $\approx$15~K for both in-phase (dispersion) and the out of phase (absorption) terms. The peak position in each curve termed as freezing temperature (T$_f$), which is generally defined as a point where thermally activated processes achieve a maximum value. We observe that the peak position shifts $\approx$2~K towards the higher temperature with increase in the frequency from 19 to 4987~Hz. For the $x=$ 0.1 sample, ac-susceptibility curves show a broad cusp below 18~K in the real and imaginary parts [see Figures~\ref{fig:ACS}(c, d)], which also shifts towards higher temperature and the magnitude of the magnetic moment decreases with the increase in the excitation frequency. This shift in the peak position of ac-susceptibility is a characteristic feature to the presence of spin glass (SG) behavior in the system \cite{SowPRB12}. The change in the magnitude of the moment indicates that the individual spin dynamics gets affected due to the excitation frequency. This cusp in the ac-susceptibility curve of $x=$ 0.1 sample, matches with the bifurcation in the temperature dependent ZFC-FC curve of dc-magnetization [see Figure~\ref{fig:MT}(c)]. 

We have performed detailed analysis of ac-susceptibility data to find the information about spin dynamics. The comparison of the change in T$_f$ with the excitation frequency is generally termed as the frequency sensitivity, also termed as the Mydosh parameter \cite{MydoshTF93} and can be calculated using the following formula $$\delta T_f = \frac{\Delta T_f }{T_f(\Delta log_{10}f)}$$ where $f$ is excitation frequency. Here we have determined the difference employing the highest and lowest measured excitation frequencies. The obtained $\delta T_f$= 0.05 and 0.04 for the $x=$ 0.05 and 0.1 samples, respectively. The values of $\delta T_f$ are generally defined in the range of 0.004--0.018 for the canonical SG behavior, and for superparamagnets (SPMs) in between 0.3-0.5 \cite{MydoshTF93,HanasakiPRL09}. We have obtained values larger than the canonical spin glasses, such as CuMn ($\delta T_f$=0.005) \cite{MydoshTF93} and smaller than the SPMs (ideal non-interacting spin glasses), like holmium borate glass a-[Ho$_2$B$_3$(B$_2$O$_3$)] \cite{MydoshTF93}. The freezing temperature dependence on the excitation frequency strongly depends on the strength of interaction betweeen the individual spins or magnetic entities. For example, stronger frequency sensitivity means the weak interaction like in magnetic clusters, while smaller value means a strong interaction between individual spins in the system. For the systems having the strong interaction (like normal ferromgnetic and antiferromganetic), large excitation frequencies (in range of MHz to GHz) are required to observe any remarkable shift in the frequency dependent ac-curves \cite{MalinowskiPRB11}. Thus, we can say that for $x=$ 0.1 sample, spin interaction is relatively stronger owing to the smaller value of $\delta T_f$ and that can be also seen in the sharp bifurcation in the dc-magnetization curve [see Figure~4(c)].

\begin{figure}
\centering
\includegraphics[width=3.4in]{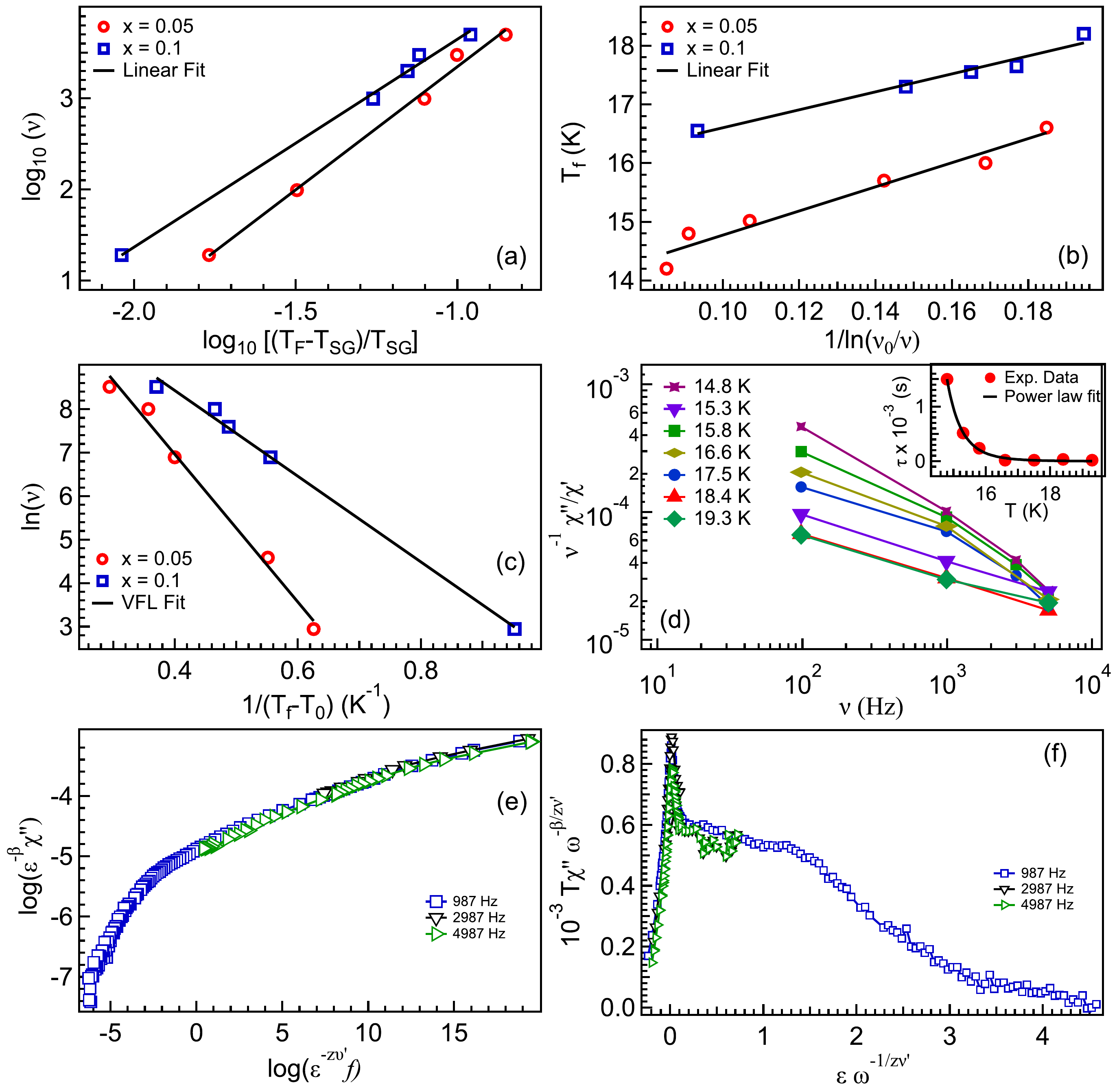}
\renewcommand{\figurename}{Figure}
\caption{(a) Frequency sensitivity of the freezing temperature is plotted in terms of log$_{10}$($\nu$) vs log$_{10}\frac{T_F-T_{SG}}{T_{SG}}$, where $\nu=1/\tau$, solid line shows the fit to the power law, (b) Excitation frequency dependence of freezing temperature is plotted as T$_f$ vs 1/ln($\nu_0$/$\nu$), solid line is fit to Vogel-Fulcher law, (c) frequency dependence of T$_f$ plotted as ln($\nu$) vs 1/(T$_f$-T$_0$) along with a fit to the Vogel Fulcher law, and (d) temperature dependent Ogielski plot for the $x=$ 0.05 sample (inset shows the power law fit of extracted frequency independent values.) (e) Critical slowing down analysis of the imaginary part of the ac--$\chi$ data at different frequencies, and (f) a linear scaling plot of $\chi''(\omega,T)/\omega^{\beta/z\nu'}$ plotted with $\epsilon/\omega^{1/z\nu'}$ using $\chi''(T)$ data, for the $x=$ 0.05 sample.}
\label{fig:VFL}   
\end{figure}

Frequency dependence of the freezing temperature T$_f$ obtained from the real part, can be analyzed using the critical slowing down of the relaxation time by applying the standard dynamical scaling theory using the relation \cite{MydoshTF93, SasakiPRB05}, $$\tau = \tau_0\left( \frac{T_f - T_{SG}}{T_{SG}}\right)^{-z\nu'}$$
where $\tau$ is the relaxation time for the measured frequency, $\tau_0$ is the characteristic relaxation time for the spin flipping or microscopic relaxation time, T$\rm_{SG}$ is the critical temperature for spin glass ordering or called static freezing temperature, i.e., when $\nu$ tends to zero, and $z\nu'$ is the dynamic exponent (in this $\nu'$ is the parameter for the correlation length $\zeta = (T_f-T_{\rm SG})^{-\nu'}$ and $\tau$ $\sim$ $\zeta^z$). Such kind of higher values have been observed elsewhere in SG and re-entrant spin glass (RSG) \cite{MydoshTF93}. For the simplicity we can write this formula in the form of $$ log_{10}(\tau) = log_{10}(\tau_0) - z\nu' log_{10}\left(\frac{T_f-T_{SG}}{T_{SG}}\right)$$ For further analysis we have plotted the data log$\rm_{10}$($\nu$) vs log$_{10}$[(T${\rm_f}$-T$\rm_{SG}$)/T$\rm_{SG}$], where $\nu=1/\tau$, in the Figure~\ref{fig:VFL} (a), which show the linear behavior for both the samples ($x=$ 0.05 and 0.1) and the values of $\tau_0$ and $z\nu'$ can be found from the intercept and slope, respectively. We obtained $\tau_0=$ 8.95$\times$10$^{-7}$ s and 1.2$\times$10$^{-6}$ s, and $z\nu'=$ 2.7 and 2.3, for the $x=$ 0.05 and 0.1 samples, respectively. It is previously reported that the conventional spin glasses have the value of $\tau_0$ in the range of 10$^{-12}$--10$^{-14}$ s and $z\nu'$ lies between 4--12 \cite{MydoshTF93, NamPRB00}. In comparison to this range $\tau_0$ is larger and that indicates towards slower spin dynamics in our system, due to the presence of strong correlation between the individual spins rather than the single non-interacting spins.

This interaction of spins can be further investigated by the Arrhenius relation; because failure of Arrhenius law confirms the presence of interaction between the individual spins as it applies for the non-interacting or weakly interacting magnetic entities. Arrhenius relation can be written as $$\nu = \nu_0 exp\left(\frac{-E_a}{k_BT_f}\right)$$ where k$_B$ is the Boltzmann constant, $\nu_0$ is the characteristic attempt frequency, and E$_a$ is the average thermal activation energy of the relaxation barrier. Arrhenius law accounts for the time scale to overcome the energy barriers by the activation process. In the plot of $ln(\nu)$ versus 1/T$_f$ (not shown), we found that the behavior is not linear towards the lower values of excitation frequencies, which also suggests the interaction of individual spins in the freezing process of the system. The fit of these values gives the unphysical outputs (E$_a$/k$_B=$ 700$\pm$75 K and 1062$\pm$190 K and $\nu_0=$ 1.7$\times$10$^{22}$~Hz and 2.6$\times$ 10$^{29}$~Hz for the $x=$ 0.05 and 0.1, respectively), that is due to the smaller number of fitting parameters and signifies the failure of Arrhenius law.

Now, to increase the number of fitting parameters, another law is more phenomenological to explore the dynamical scaling behavior of spin freezing is the Vogel-Fulcher(VF) law defined with an empirical relation $$\nu = \nu_0 exp\left(\frac{-E_a}{k_B(T_f-T_0)}\right)$$ where $T_0$ is the characteristic temperature or Vogel-Fulcher temperature used to measure interaction strength between magnetic entities, $\nu_0$, and E$_a$ have the same physical meaning as defined previously. We fitted our data taking two methods first via fixing the previosly obtained value of $\nu_0$, and in order to fit the data it is convenient to write the equation in terms of $$ln\left(\frac{\nu_0}{\nu}\right) = \frac{E_a/k_B}{(T_f-T_0)}$$ and that can be rearranged to $$T_f = \frac{E_a/k_B}{ln(\nu_0/\nu)} + T_0$$A plot between T$_f$ vs 1/ln$({\nu_0}/{\nu})$ allows us to determine the E$\rm_a$/k$\rm_B$ and $\nu_0$ from the slope and intercept, respectively [see Figure~\ref{fig:VFL}(b)]. Fit results give the values of E$_a$/k$_B$ = 20.5~K and 15.3~K, and T$_0$ = 12.7~K and 15.1~K, for the $x=$ 0.05 and 0.1 samples, respectively. In order to make sure these output fitting parameters as a result of fixing the $\nu_0$ parameter are correct, we have also fitted the data fixing the T$_0$ parameters. This Vogel-Fulcher temperature (T$_0)$ is determined by the method suggested by Souletie and Tholence \cite{SouletiePRB85} and obtained 13.5$\pm$0.3 and 16$\pm$0.2~K, for $x=$ 0.05 and 0.1, respectively. We used simplified relation $$ln(\nu) = ln(\nu_0) - \frac{E_a/k_B}{T_f-T_0}$$ and plotted a curve between ln$(\nu)$ versus 1/($T_f-T_0$), where the best fit of data give value of characteristic relaxation time $\tau_0$ = 1.1$\times$10$^{-6}$ and 4.2$\times$10$^{-6}$ s, and activation energy E$_a$/k$_B$ = 17$\pm$1 and 10$\pm$0.5~K for the $x=$ 0.05 and 0.1 samples, respectively [see Figure~\ref{fig:VFL}(c)]. From the above analysis, we determine that T$_0$ is nonzero for our system and that arises because of the interaction between the individual spins, which can be considered as a behavior of cluster-spin glass \cite{GunnarssonPRL88, AnandPRB12, MukadamPRB05}. As we found that the value of T$_0$ is close to the freezing temperature, suggests the presence of Ruderman-Kittel-Kasuya-Yosida (RKKY) interaction. Using the Tholence criterion \cite{TholencePB84} $\delta T_{Th} = (T_f-T_0)/T_f$, where T$_f$ is measured at smallest excitation frequency, we obtained $\delta T_{Th}$ = 0.07 and 0.06 for the $x=$ 0.05 and 0.1 samples, respectively. These values are comparable to the RKKY spin glass systems (e.g., 0.07 for the CuMn system) \cite{TholencePB84}, and that suggest that our system falls in the category of RKKY spin glass. It is important to note here that the RKKY interactions are normally present in metallic systems and cannot present in band insulators. However, Hellman {\it et al.} reported spin-glass freezing with mixed antiferromagnetic and ferromagnetic interactions, and RKKY interaction in amorphous Gd-Si alloys, which are insulating or bad metallic, but nor band insulator \cite{HellmanPRL00, HelgrenPRB07}. The nature of RKKY-type indirect exchange mediated by high density of electrons, localized due to disorder where the variation of the density of states N(E) across the metal insulator transition may play crucial role \cite{HellmanPRL00, BokachevaPRB04}. Interestingly, Androulakis {\it et al.} proposed that a few existing itinerant electrons in LaCoO$_3$ can couple ferromagnetically  via short range RKKY interactions at low temperatures \cite{AndroulakisPRB01}. In this context note that the La$_{(1-2x)}$Sr$_{2x}$Co$_{(1-x)}$Nb$_{x}$O$_3$ samples show insulating nature, but not a band insulator \cite{AndroulakisPRB01, RadwanskiPB05} and it is consistent for the $x=$ 0.05 sample where we observed larger value of N(E) from resistivity analysis, discussed later. This analysis motivates for further experimental and theoretical studies on LaCoO$_3$.

We found that the value of $\tau_0$ obtained from the power law is one order of magnitude smaller than obtained from the VF law for the $x=$ 0.05, whereas for the $x=$ 0.1 both values are comparable. Further, in the frame of VF model, when T$_0$ $>$ E$a$/k$_B$ indicates a strong coupling, while T$_0$ $<$ E$a$/k$_B$ signifies the weak coupling between magnetic entities \cite{ShtrikmanPLA81}. For our case as observed prior $x=$ 0.1 has strong coupling in comparison to $x=$ 0.05, sample. To further understand the spin glass dynamics we analyzed the data according to the formalism proposed by Ogielski \cite{OgielskiPRB8587}, $\lim_{\nu\to0}$ $\chi''(\nu)$/$\chi'(\nu)$ = $\tau_{av}$, where $\tau_{av}$ is the average correlation time. We have plotted $\nu^{-1}(\chi''/\chi')$ vs $\nu$ in the log-log scale near the T$_f$ in the Figure~\ref{fig:VFL}(d). At a given value of T, as depicted above we get a frequency independent value of $\nu^{-1}(\chi''/\chi')$ in lower frequency limit. This frequency independent value is estimated by extrapolating each curve up to the lowest excitation frequency ($\nu=$ 9~Hz). Fitting the values extracted from the low frequency limit results the $\tau_0$=1.5$\times$10$^{-7}$ s, T$_{\rm SG}=$11.5~K. The value of $\tau_0$ for the $x=$ 0.05 sample is comparable to the calculated previously, which confirms that the spin dynamics is slower than expected for the typical SG systems. 

For the dynamical scaling of the imaginary part of ac-susceptibility, two exponents z$\nu'$ and $\beta$ are used and we followed the scaling formulation given below \cite{GeschwindPRB90,MarcanoPRB19}, $$\chi''(f,T) = \epsilon^\beta F(f\epsilon^{z\nu'})$$ where F is the universal scaling function, $\epsilon$ = (T-T$_{SG}$)/T$_{SG}$, and $\beta$ is the order parameter. In Figure~\ref{fig:VFL}(e), we present the dynamical scaling for $x$=0.05 sample, where all the curves fall on the same line with the values of z$\nu'=$ 8.5 and $\beta=$0.5. For the further verification of these values with the theory, we have also plotted in Figure ~\ref{fig:VFL}(f) the other scaling equation \cite{BitlaPRB12,NairPRB07} in terms of $$T \chi''(\omega,T)/\omega^{\beta/z\nu'} = G(\epsilon/\omega^{1/z\nu'})$$ where G is the scaling function, which is expected to highlight any departures from the scaling equation due to higher sensitivity of the abscissa and ordinate scales, and it resembles the plot of $\chi''$ versus temperature curve for all frequencies merged to a single curve, see Figure~8(f).

\subsection{\noindent ~D. Low temperature neutron study}

Now for the detailed investigation of magnetic phase, we have performed neutron diffraction at 3~K, see Figures~9(a--c). 
\begin{figure}[h]
\centering
\includegraphics[width=3.4in]{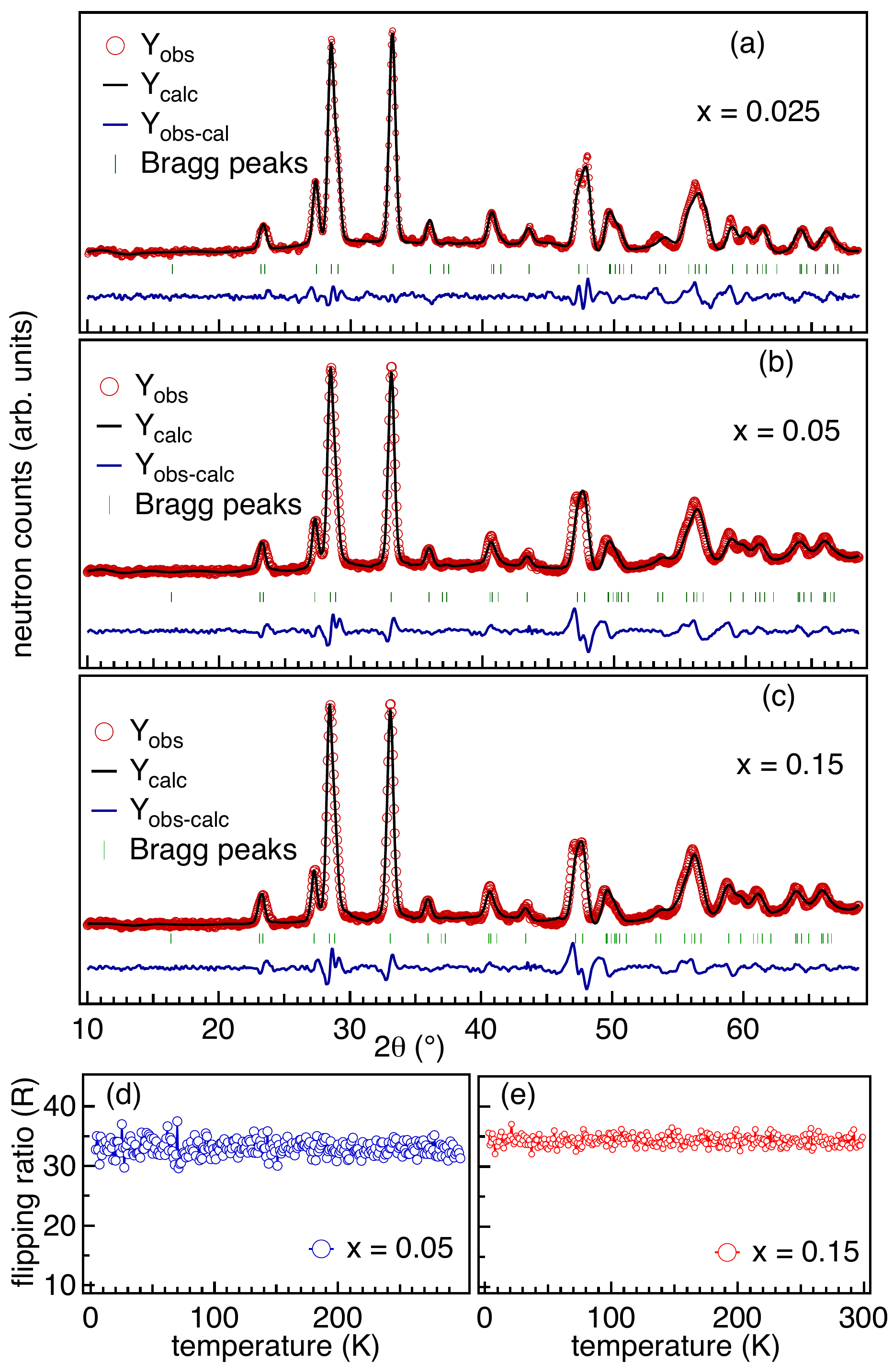}
\renewcommand{\figurename}{Figure}
\caption{The Rietveld refined neutron powder diffraction data for the La$_{(1-2x)}$Sr$_{2x}$Co$_{(1-x)}$Nb$_{x}$O$_3$ samples, (a) $x=$ 0.025, (b) $x=$ 0.05, and (c) $x=$ 0.15, measured at 3~K. (d) and (e) show the temperature variation of the flipping ratio (R) for $x=$ 0.05 and 0.15, respectively.}
\label{fig:DP}
\end{figure}
Neither additional magnetic Bragg peaks nor enhancement in the intensity of fundamental nuclear Bragg peaks has been observed at 3~K \cite{DurandJPCM13}, ruling out the presence of a ferromagnetic/antiferromagnetic ordering \cite{SikolenkoJPCM09,RajeevanJMMM15}. Further, we carried out neutron depolarization study down to 3~K on the $x=$ 0.05 and 0.15 samples [Figures~9(d, e)], where no depolarization of neutron beam was observed ruling out the presence of any FM domains/clusters.

\subsection{\noindent ~E. Resistivity measurements}

\begin{figure}[h]
\centering
\includegraphics[width=3.6in]{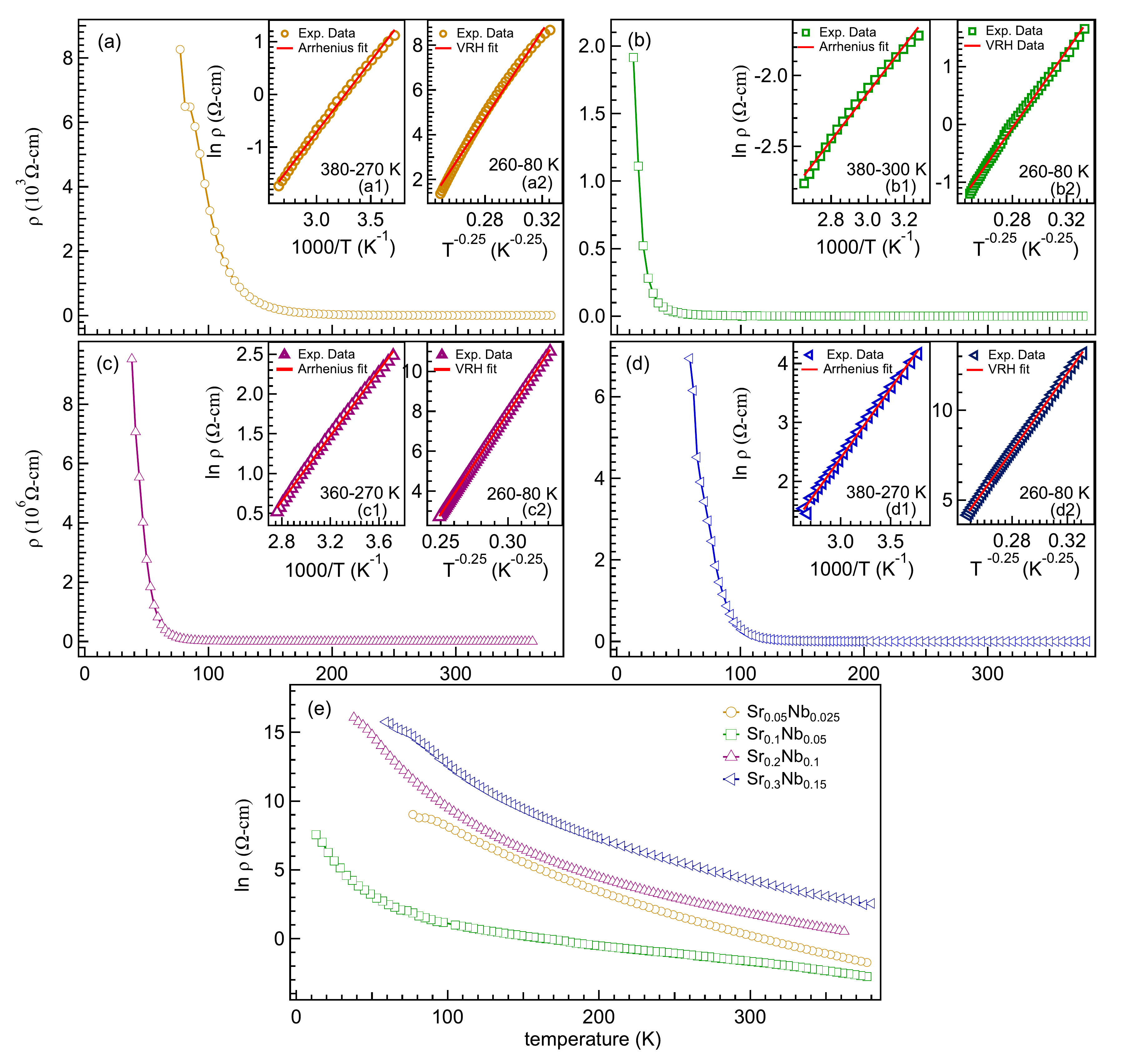}
\renewcommand{\figurename}{Figure}
\caption{Temperature dependent resistivity data for the La$_{(1-2x)}$Sr$_{2x}$Co$_{(1-x)}$Nb$_{x}$O$_3$ samples (a) $x=$ 0.025, (b) $x=$ 0.05, (c) $x=$ 0.1, and (d) $x=$ 0.15, inset shows the fitting in different temperature ranges. (e) the comparison of resistivity values for all the samples in logarithmic scale.}
\label{fig:Res}
\end{figure}

The parent sample LaCoO$_3$ shows an insulating ground state, which can be tuned towards the metallic state by substituting Sr at La site \cite{WuPRB03}, whereas the substitution of Nb at Co site drives the system towards the insulating regime \cite{OygardenJSSC12}. Here we study the effect of co-substitution on the transport properties. The resistivity data are presented in Figure~\ref{fig:Res}. 
The electrical resistivity data show that all the samples are insulating and the resistivity increases from $\ohm\cdot$cm to M$\ohm\cdot$cm when the temperature decreases from 380~K to low temperatures. The electrical resistivity at low temperatures could not be measured because of compliance limit of the measuring instrument and highly insulating nature of $x=$ 0.025 and 0.15 samples. To explain the electrical conduction we have fitted the data with Arrhenius and variable range hopping (VRH) models in the high and low temperature regimes, respectively \cite{ShuklaPRB18}. We use the localization length comparable to the La/Co--O bond length and estimated the value of density of states near the Fermi level N(E$_F$)(eV$^{-1}$cm$^{-3}$) using the VRH model. 
\begin{table}[h]
		\centering
		\label{tab:RT}
		\renewcommand{\tablename}{Table}
		\renewcommand{\thetable}{\arabic{table}}
		\caption{Calculated values of density of states N(E) by VRH model fitting and activation energy E$_a$ by Arrhenius equation fitting, and other fitting parameters for La$_{(1-2x)}$Sr$_{2x}$Co$_{(1-x)}$Nb$_{x}$O$_3$ samples.}
		\vskip 0.2 cm
		\begin{tabular}{|c|c|c|}
		\hline
		\textbf{samples}& VRH Model & Arrhenius Model\\
		$x$(\%)&T$_0$(K), \hskip 0.4cm N(E)(eV$^{-1}$cm$^{-3}$)& E$_a$(meV) \\
		\hline 
		0.025&8.9$\times$10$^7$, \hskip 0.7cm 3.2$\times$10$^{20}$ &  144$\pm$3 \\
		\hline
		0.05&1.2$\times$10$^6$, \hskip 0.7cm 2.3$\times$10$^{22}$ &  132$\pm$2 \\
		\hline
	         0.1&1.1$\times$10$^8$, \hskip 0.7cm 2.7$\times$10$^{20}$ &  177$\pm$2 \\
		\hline
	        0.15&1.4$\times$10$^8$, \hskip 0.7cm 2.0$\times$10$^{20}$&  204$\pm$2 \\
	\hline	
		\end{tabular}
\end{table}
The density of states decreases with $x$, this suggest that the samples are becoming more insulating in nature. The calculated values of the activation energy show that the insulating nature enhances with the increased substitution, and is in meV range [see Table~IV], however, such small activation energy is expected to result in a semiconducting nature as it is in the case of LaCoO$_3$ with E$_a$$\approx$120 meV \cite{ZobelPRB02}. We have calculated the value of activation energy as well as density of states near the Fermi level using the fitting parameters, as presented in the Table~IV. Interestingly, for the $x=$ 0.05 sample, the activation energy is significantly lower and N(E$_{\rm F}$) is also two order of magnitude higher than the other samples \cite{EnglishPRB02}. This makes it consistent with the observed signature of RKKY like interaction in this sample from ac-susceptibility analysis, as discussed earlier.

\subsection{\noindent ~F. X-ray photoelectron spectroscopy study}

\begin{figure}[h]
\centering
\includegraphics[width=3.3in]{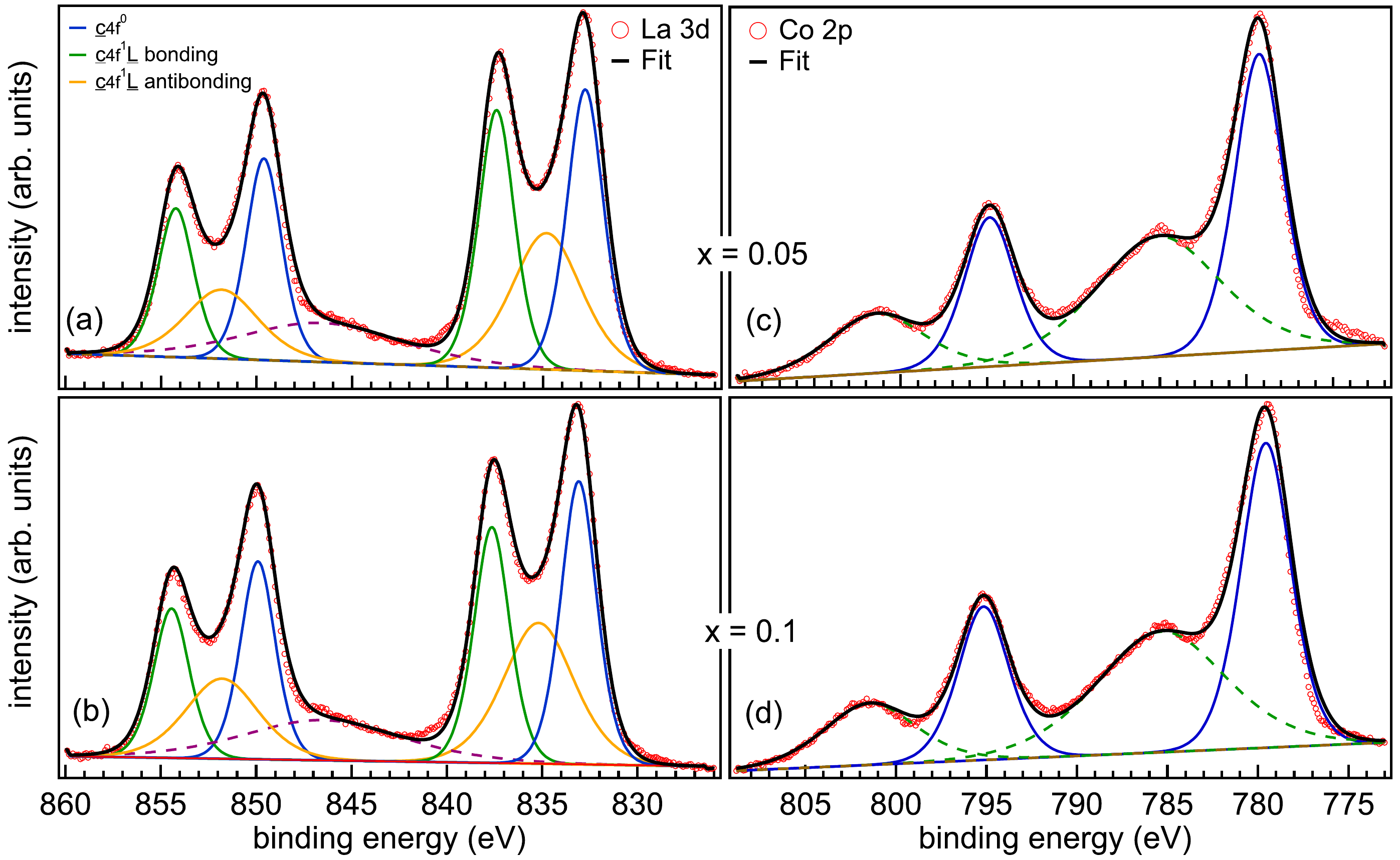}
\renewcommand{\figurename}{Figure}
\caption {The La 3$d$ (a, b) and Co 2$p$ (c, d) core level photoemission spectra of La$_{(1-2x)}$Sr$_{2x}$Co$_{(1-x)}$Nb$_{x}$O$_3$ ($x=$ 0.05 and 0.1) samples along with fitting components.}
\label{fig:XPS_LC}
\end{figure}

\begin{figure}[h]
\centering
\includegraphics[width=3.3in]{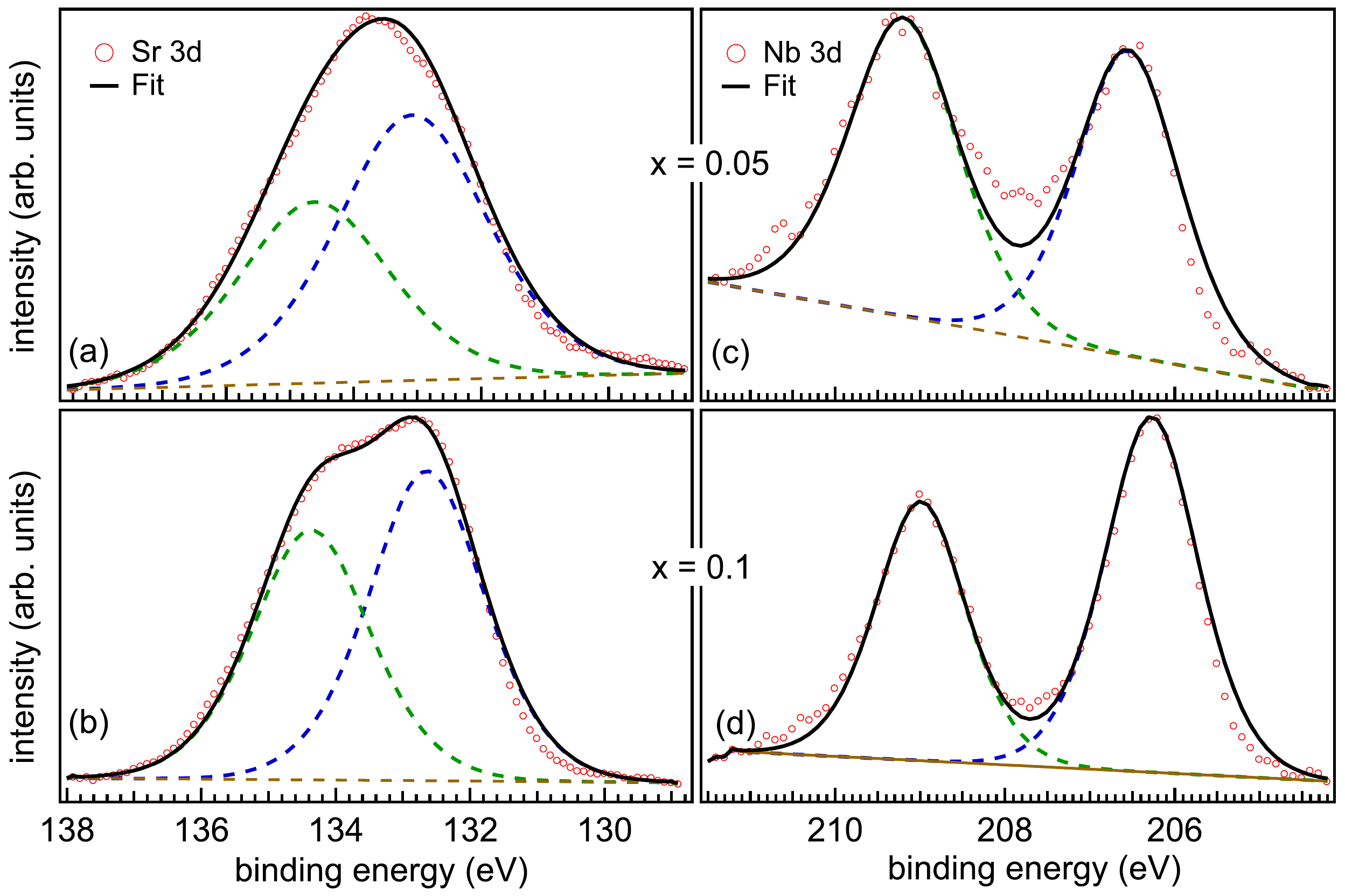}
\renewcommand{\figurename}{Figure}
\caption {The core level photoemission spectra of substituted elements, Sr 3$d$ (a, b) and Nb 3$d$ (c, d) in La$_{(1-2x)}$Sr$_{2x}$Co$_{(1-x)}$Nb$_{x}$O$_3$ ($x=$ 0.05 and 0.1) samples along with fitting components.}
\label{fig:XPS_SN}
\end{figure}

We now move to the study of electronic properties of $x=$ 0.05 and 0.1 samples using x-ray photomeission spectroscopy (XPS). The room temperature XPS core level spectra are presented in Figures~\ref{fig:XPS_LC} and \ref{fig:XPS_SN}. We observe that there is no significant shift in peak position within the limit of instrumental resolution. The Co 2$p$ core level spectrum shows two main peaks, (corresponds to 2$p_{3/2}$ and 2$p_{1/2}$ at 779.5~eV and 795~eV, respectively) owing to the spin-orbit splitting of $\sim$15.5~eV and an intensity ratio close to 2:1 \cite{ChainaniPRB92, ChuangSS76}. These observations confirm that the Co is predominantly in the Co$^{3+}$ state, as reported for LiCoO$_2$ \cite{MosesASS07} and Na$_x$CoO$_2$ \cite{JugoviJALCOM19}. It also shows two loss features for each peak towards the higher binding energy (BE) side, known as the shake up satellites. The position of this satellite may suggest the presence of Co$^{2+}$; however, if there is even 5\% Co$^{2+}$ present in the sample, the $\mu_{\rm eff}$ increases from $\approx$3.85 to 4.65~$\mu_{\rm B}$ \cite{ShuklaPRB18}, which is significantly larger than what is observed from the magnetization measurements up to $x=$ 0.1 sample, see table~3. Therefore, the formalism that the Co$^{3+}$ valence state is preserved in LaCoO$_3$, by co-substitution of Sr and Nb in 2:1 ratio, is consistent. In the transition metal oxides the satellite features can be explained with the charge transfer processes using molecular orbital theory or sudden approximation \cite{HufnerBOOK}. The ground state of Co$^{3+}$ can be associated with a ligand (O$^{2-}$ for the oxides) like 3d$^6$L, where L is the highest occupied ligand shell i.e., 2$p$ for the O$^{2-}$. After the photoexcitation $d$ band lowers its energy and an electron is transferred from the ligand and that may lead to the satellite feature in Co 2$p$ core-level.  

The La 3$d$ spectrum shows four peaks \cite{ShuklaPRB18}, two peaks for each spin orbit component 3$d_{5/2}$ and 3$d_{3/2}$ alongwith two extra peaks due to the transfer of an electron from the oxygen valence band to the empty La 4$f^0$ level \cite{LamPRB80}. The ionised La atoms have the 3$d$ core level and 4$f$ is the nearest unoccupied orbital, so a charge transfer from the ligand valence band O(2$p$) to the 4$f^0$ orbital and transfer of an electron from the 3$d$ core level to the continuum is believed a reason behind this complex structure \cite{MullicaPRB85, SignorelliPRB73}. When we remove an electron from the core level we make the La ions more electronegative and then there is a possibility of electron withdrawal from the oxygen anions. Intensity of the observed satellite peaks depend on the electronegativity of the ligand molecule. For the fitting of La 3$d$ core level spectra we follow the formalism used by Mullica {\it et al.} and hence each spin-orbit component is fitted with three components \cite{MullicaPRB85}. The main peaks at the lower binding energy (832.8~eV and 849.6~eV for the 3d$_{5/2}$ and 3d$_{3/2}$, respectively) corresponds to the final state without charge transfer, i.e., 3$d^9$4$f^0$ in La$^{3+}$ and denoted by the \underline{c}4$f^0$, where \underline{c} denotes the presence of a core hole and 4$f^0$ indicates absence of an electron in the 4$f$ orbital. Other two peaks at the higher binding energies are bonding and antibonding of the final state denoted as \underline{c}4$f^1$\underline{L}, which indicates a charge transfer from O(2$p$) ligand to the metal 4$f$ orbital. The respective de-convoluted components are defined in Figure~11(a). It is important to note that unoccupied 4$f$ orbital of the La$^{3+}$ resides above the valence band; therefore an electron transfer to the 4$f$ orbital will be an energy loss process \cite{SignorelliPRB73, SarmaCP82}. Figures~\ref{fig:XPS_SN} (a, b) show Sr 3$d$ and Nb 3$d$ core levels, which are fitted using Voigt function in CasaXPS software, after subtracting inelastic Tougaard background. The position of Sr 3$d_{5/2}$ core level has been observed at 132.8~eV binding energy (BE) corresponds to the valency of 2+, which is close to the reported in refs.~\cite{JongPRB09,GuoPRB10}. The Nb 3$d$ core levels of both the samples are shown in Figures~\ref{fig:XPS_SN} (c, d), which clearly have spin orbit peaks 3$d_{5/2}$ and 3$d_{3/2}$ at 206.5~eV and 209~eV, respectively. This confirms the existence of Nb in the 5+ valence state \cite{ShuklaPRB18, SasaharaJPCC13}.

\section*{\noindent ~Conclusions}

In this study we have investigated the structural, magnetic, and electronic transport properties with the Sr/Nb co-substitution (hole as well as electron) in LaCoO$_3$, i.e., La$_{(1-2x)}$Sr$_{2x}$Co$_{(1-x)}$Nb$_{x}$O$_3$. The recorded room temperature powder x-ray diffraction data are refined with rhombohedral symmetry, and no structural transition has been observed, but unit cell volume increases with $x$. The magnetization values increases significantly with respect to the $x=$ 0 sample and the coercivity also changes. The calculated values of effective magnetic moment confirm the spin state of Co$^{3+}$ in intermediate and high spin-states in the 50:50 ratio. The ac-susceptibility measurements show the change in T$_f$ with the excitation frequency, which indicate spin-glass behavior in the system. The extracted relaxation time suggests slower spin dynamics, which confirms the cluster spin glass. Also, the Tholence criterion suggests the presence of RKKY interactions in $x=$ 0.05 sample, which is consistent as the density of states are two order of magnitude larger in this case. The neutron diffraction and depolarization data provide a strong support at microscopic and mesoscopic length scales, respectively, to the observed spin-glass state.The dc-transport properties manifest that co-substitution drives the system towards insulating regime, and we extracted the activation energy and density of states. The x-ray photoemission measurements confirm that the Co ions predominantly exist in the 3+ valence state and shakeup satellite features exists due to the ligand interaction. The La 3$d$ core level spectra show a complex peak structure due to the electron transfer between the ligand and unoccupied 4$f$ level. We also confirm the valence state of Sr and Nb as 2+ and 5+, respectively.

\section*{\noindent ~Acknowledgments}

RS gratefully acknowledge the DST-Inspire, India for fellowship. We would like to thank the research facilities: XRD, PPMS EVERCOOL-II at IIT Delhi. The powder neutron diffraction and ac-susceptibility measurements are supported under CRS project no. UDCSR/MUM/AO/CRS-M-270/2017/567. We thank Ploybussara Gomasang for help in the XPS measurements. The XPS measurements were supported by Sakura Science Program (aPBL), Shibaura Institute of Technology, under the Top Global University Project, Designed by Ministry of Education, Culture, Sports, Science and Technology in Japan. We also thank Vivek Anand, Priyanka, Mahesh Chandra, and Ajay Kumar for useful discussions and help. We thank SERB-DST for financial support through Early Career Research (ECR) Award (project reference no. ECR/2015/000159). Some facilities have been used in this study which are supported by BRNS through DAE Young Scientist Research Award project sanction No. 34/20/12/2015/BRNS.

\end{document}